\documentclass[review,12pt]{elsarticle}
\usepackage[margin=1in,footskip=0.25in]{geometry}
\usepackage{amssymb,amsmath,latexsym,amsthm,amsfonts,mathtools}

\usepackage{mathrsfs}
\usepackage{multirow}
\usepackage{graphicx}
\usepackage{textcomp,siunitx}
\usepackage{float}
\usepackage{subcaption}
\usepackage{booktabs}
\usepackage{enumitem}

\usepackage{bm}
\usepackage{hyperref}
\hypersetup{colorlinks, breaklinks, citecolor=blue, linkcolor=red, urlcolor=blue}
\usepackage[nameinlink,noabbrev,sort&compress]{cleveref}
\usepackage{romannum}

\usepackage{siunitx}
\usepackage{longtable,tabularx}
\usepackage{booktabs}
\usepackage[dvipsnames]{xcolor}
\usepackage{tikz}
\usetikzlibrary{positioning}
\usetikzlibrary{shapes,arrows,automata}
\usepackage{wrapfig}
\allowdisplaybreaks

\theoremstyle{plain}
\newtheorem{theorem}{Theorem}
\newtheorem{lemma}{Lemma}
\newtheorem{corollary}{Corollary}

\theoremstyle{remark}
\newtheorem{remark}{Remark}

\theoremstyle{definition}
\newtheorem{definition}{Definition}

\newtheorem{problem}{Problem}

\newcommand{\sign}{\mathrm{sign}}

\usepackage{cleveref}
\Crefformat{figure}{#2Fig.~#1#3}
\Crefmultiformat{figure}{Figs.#2#1#3}{ and#2#1#3}{, #2#1#3}{ and~#2#1#3}
\Crefformat{equation}{#2Eq.~#1#3}
\Crefmultiformat{equation}{#2Eqs.#1#3}{ and#2#1#3}{, #2#1#3}{ and~#2#1#3}
\allowdisplaybreaks

\usepackage{lineno}
\modulolinenumbers[5]

\journal{Aerospace Science and Technology}

\biboptions{numbers,sort&compress}

\makeatletter
\def\ps@pprintTitle{%
	\let\@oddhead\@empty
	\let\@evenhead\@empty
	\let\@oddfoot\@empty
	\let\@evenfoot\@empty
}
\makeatother

\begin{document}
\pagenumbering{arabic}

\begin{frontmatter}

\title{Field-of-View and Input Constrained Impact Time Guidance Against Stationary Targets}
\author[srk]{Swati Singh}
\corref{mycorrespondingauthor}
\cortext[mycorrespondingauthor]{Corresponding author}
\ead{swatisingh@aero.iitb.ac.in}

\author[srk]{Shashi Ranjan Kumar}
\ead{srk@aero.iitb.ac.in}
\address[srk]{Intelligent Systems and Control Lab, Department of Aerospace Engineering, Indian Institute of Technology Bombay, Powai, Mumbai - 400076, India}
\author[dm]{Dwaipayan Mukherjee}
\ead{dm@ee.iitb.ac.in}
\address[dm]{Department of Electrical Engineering, Indian Institute of Technology Bombay, Powai-- 400 076, Mumbai, India}
	
\begin{abstract} 
This paper proposes a guidance strategy to achieve time-constrained interception of stationary targets, taking into account both the bounded field-of-view (FOV) of seeker-equipped interceptors and the actuator's physical constraints. Actuator saturation presents a significant challenge in real-world systems, often resulting in degraded performance. However, since these limitations are typically known in advance, incorporating them into the guidance design can enhance overall performance. To address the FOV constraint, a time-to-go error-based approach is adopted. Furthermore, to incorporate the lateral acceleration constraints, the engagement kinematics are augmented with an input saturation model. Subsequently, the guidance strategy that constrains the lateral acceleration and the time-to-go values within their respective bounds is derived using Lyapunov stability concepts and the backstepping technique. Furthermore, a multi-stage approach is suggested to expand the achievable range of impact time. Numerical simulations are performed to validate the efficacy of the proposed scheme for different initial engagement geometries.
\end{abstract}
\begin{keyword}
    Impact time, Field-of-view constraints, Nonlinear guidance, Input saturation
\end{keyword}
\end{frontmatter}

\section{Introduction}
Most classical guidance laws aim to achieve precise target interception. However, a single interceptor may not be sufficient against advanced adversaries with defensive systems like close-in weapons. Coordinated salvo attacks are more effective in such cases, as they increase the likelihood of neutralizing the threat. These attacks require precise control over the interception time of each interceptor, which presents a significant challenge. Achieving this synchronization enables the salvo to breach the adversary’s defenses, motivating research into guidance strategies that ensure interception at a specific, predefined impact time \cite{1597196,doi:10.2514/1.G001681,doi:10.2514/1.G001618}.

Impact time-constrained target interception is a well-explored problem in the literature, addressed through various approaches. Impact time guidance algorithms have been developed using methods such as Lyapunov theory, sliding mode control (SMC), and optimal control \cite{jeon2006impact, jeon2016impact, cho2015nonsingular}. In \cite{jeon2006impact}, the problem was formulated as error regulation with linearized kinematics, and an optimal control strategy based on proportional navigation guidance (PNG) was designed to minimize the impact time error. This was extended in \cite{jeon2016impact} within a nonlinear framework. In \cite{cho2015nonsingular}, an exact analytical solution for time-to-go was provided using PNG law and nonlinear engagement kinematics, which was globally accurate for all flight directions. A composite guidance law with a two-phase variation for the lead angle was presented in \cite{Cheng2018comp_AST}. In the first phase, the lead angle at the launch is maintained, and in the second phase, the lead angle linearly decreases to zero at target interception. Under this strategy, the interceptor does not take the collision course. It is, conversely, more advantageous to track the collision course as it maximizes the damage inflicted.
A feedback controller over a biased PNG law was introduced in \cite{Chen2019OGL_AST} to drive the impact time of the nominal trajectory to the desired value.
Another PNG-based feedback controller was discussed in \cite{CHEN2019454} for precise interception. A two-stage impact time control approach was introduced in \cite{doi:10.2514/1.G001719}, though the radial acceleration term therein may pose implementation challenges. 
In \cite{Harl2012ITACG}, the line-of-sight (LOS) angle was shaped as a polynomial function to satisfy both impact time and impact angle constraints, with a 90$^\circ$ flight path angle bound. Impact time control necessitates that time-to-go, a quantity associated with flight conditions in the future, be known a priori. However, obtaining a closed-form
expression for the same is not a trivial task \cite{ZHU2019818}. Integrated guidance and control design was proposed in \cite{GUO201854,MING2019105368} to directly generate the fin deflections, significantly enhancing the endgame performance of the homing interceptor.
Unlike the previous work, in addition to the guidance law, the autopilot was also designed in \cite{Sinha2021nested_AST} for a dual-controlled interceptor, having canard as well as tail control inputs. For a similar interceptor configuration, an integrated guidance and control (IGC) strategy was presented in \cite{Sinha2021integ_AST}, owing to the effectiveness of IGC over the separate guidance and control design. Authors in \cite{XIE2024109111} proposed a terminal guidance law with impact time and angle constraints, assuming controllable missile speed, simplifying time-to-go adjustment against maneuvering targets. An advanced terminal guidance algorithm designed to simultaneously satisfy impact time, impact angle, and terminal load factor constraints was designed in \cite{ZHANG2022107462}.
Furthermore, the guidance strategy \cite{ZHANG2022107462} worked even when the missile's speed was not constant and varied passively and could not be actively controlled. 
While the guidance laws in \cite{jeon2006impact, jeon2016impact, CHEN2019454, cho2015nonsingular,doi:10.2514/1.G001719,XIE2024109111} successfully achieved target interception at the desired impact time, they did not consider the limited field-of-view of the seeker. Incorporating this consideration makes the guidance law more practical, as the seeker on the interceptor has a restricted field-of-view. 

In practical scenarios, guidance designs must address the interceptor's limited field-of-view (FOV), especially for impact time-constrained laws. While achieving a desired impact time, the interceptor may need to follow a trajectory that occludes the target from the seeker. Although several impact time control laws exist, few consider FOV constraints \cite{zhang2014impact, kim2018backstepping, jeon2017impact, kim2019look}. For example, \cite{zhang2014impact} developed an impact time control guidance (ITCG) law combining PNG with impact time error for planar engagements, ensuring FOV-constrained interception. In \cite{kim2018backstepping}, a backstepping-based approach was used to define homing and impact time errors, with the homing error converging when the flight path angle matched the LOS angle. \cite{jeon2017impact} introduced PNG with a time-varying navigation gain, ensuring monotonic convergence of the lead angle, but limiting the maximum achievable impact time. \cite{kim2019look} proposed an FOV-constrained strategy that successfully achieves both the desired impact time and angle.
A terminal sliding-mode control (TSMC) impact time guidance law incorporating an FOV constraint was proposed in \cite{chen2018nonsingular}. This law, derived using a nonlinear model, eliminates the need for time-to-go estimation by designing a switching surface. The TSMC ensured that the switching surface was reached within a finite time before interception. In \cite{kumar2021three}, a barrier Lyapunov-based guidance strategy using time-to-go estimates was developed to ensure interception at the desired impact time. This strategy remained nonsingular for the achievable range of impact times and was applicable to both planar and three-dimensional engagements while considering FOV constraints.
Authors in \cite{LI2024109543} proposed a generalized circular impact time guidance.

Note that although the guidance strategies discussed above ensure target interception at a desired impact time while satisfying FOV constraints, they do not account for input constraints. Physical systems are limited by actuator capabilities, so considering input constraints is essential for real-world applicability. A field-of-view and input-constrained impact time guidance strategy was proposed in \cite{doi:10.2514/1.G007770}, where asymmetric time-varying barrier Lyapunov functions were used to bound both FOV and lateral acceleration. However, the exact bound on lateral acceleration was not specified. Additionally, the guidance strategy given in \cite{doi:10.2514/1.G007770} could not guarantee boundedness of lateral acceleration at larger impact time without the use of the ad-hoc saturation block. Building on these results, the main contribution of this paper may now be summarized below$\colon$

Firstly, a novel nonlinear guidance strategy is proposed that meets impact time requirements while respecting the interceptor’s seeker field-of-view and actuator limitations.
Secondly, a backstepping-based approach is used to derive guidance commands for nonlinear engagement scenarios, ensuring efficient operation even with large initial heading errors.
Thirdly, to address input constraints, an input saturation model is incorporated. Unlike \cite{doi:10.2514/1.G007770}, this approach provides an exact bound for the interceptor's lateral acceleration values.  
As a fourth contribution, the proposed design guarantees convergence of the interceptor's lateral acceleration and lead angle (heading error) to near-zero at the time of interception.
Fifth, unlike many existing strategies \cite{chen2018nonsingular, kumar2021three, zhang2014impact}, the proposed method does not require a saturation block to restrict lateral acceleration. It also offers a stronger stability guarantee by keeping system states within their specified bounds. Finally, a multi-stage approach is introduced to achieve interception at larger impact time values.

\section{Problem Formulation}\label{sec2}
\begin{figure}[!ht]
\centering
\includegraphics[width=0.35\linewidth]{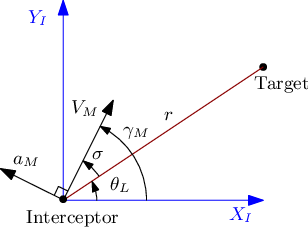}
    \caption{A typical interceptor-target engagement for the planar case.}
    \label{fig:1}
\end{figure}

Consider an interceptor and a target engagement scenario in a planar setting, as shown in \Cref{fig:1}. It is assumed that the interceptor is initially located at the origin of the inertial frame of reference represented by two mutually orthogonal axes $X_I$ and $Y_I$, with the origin of the inertial frame initially coinciding with the interceptor's center of gravity. Let $r$ and $\theta_L$ represent the relative separation and the line-of-sight angle between the interceptor and the target, $\gamma_M$ denote the heading angle, and $\sigma$ denote the velocity lead angle. The interceptor is assumed to be moving with constant speed, $V_M$. The term $a_M$ represents the lateral acceleration of the interceptor, which acts perpendicular to the velocity vector, and serves as the control input to the interceptor.
The engagement kinematics governing the relative motion between the interceptor and the target in \Cref{fig:1} are given by
\begin{equation}\label{eq:1a1b1c}
	\dot{r} = -V_M \cos\sigma,\quad
	\dot{\theta}_L =-\frac{V_M \sin\sigma}{r},\quad
	\dot{\sigma} =\frac{a_{M}}{V_M}-\dot{\theta}_L.
\end{equation}
The expression for $\dot{r}$ describes the rate at which the relative separation between the interceptor and the target changes. Similarly, $\dot{\theta}_L$ governs the rate at which the line-of-sight between the interceptor and the target rotates, while $\dot{\sigma}$ relates the interceptor's heading angle to the control input, $a_M$. By analyzing the dynamics of the interceptor's heading angle in $\dot{\sigma}$, it can be inferred that the heading angle exhibits a relative degree of one with respect to the interceptor's lateral acceleration, $a_M$. 

We now present two key temporal quantities for impact time-constrained guidance, defined as follows.
\begin{definition}
At any point during the engagement, the \emph{time-to-go}, denoted by $t_{\rm go}$, refers to the total remaining time before the target is intercepted, \cite{zarchan2012tactical}. 
\end{definition}
\begin{definition}
For an interceptor-target engagement, \emph{impact time}, $ t_{\rm imp}$, is defined as the total time taken by the interceptor to intercept the target, starting from its launch instant or any pre-specified reference \cite{kumar2021three}. Mathematically, one may write $ t_{\rm imp}= t_{\rm el} + t_{\rm go}$, where $t_{\rm el}$ represents the time that has passed since the interceptor was launched, and $t_{\rm go}$ is the time-to-go, that is, the duration remaining until the target interception takes place.
\end{definition}
In planar engagements, the field-of-view restrictions can be viewed as a bound on $\sigma$, ensuring that $|\sigma|$ satisfies the constraint $|\sigma| \leq \sigma_{\max} < \pi/2$, where $\sigma_{\max}$ represents the maximum allowable value of $\sigma$. As a result of this assumption, the target can be confined within a cone centered at the interceptor, with a half-angle of $\sigma_{\max}$, in order to stay within the interceptor's field-of-view bounds.
We are now ready to formally state the main problem addressed in this paper.
\begin{problem}\label{prob}
Given the initial conditions of the interceptor, including its heading angle $\gamma_M$, constant speed $V_M$, initial range to the target $r$, the field-of-view (FOV) bound $\sigma_{\max}$, and the maximum bound on the lateral acceleration $a_{\max}$, the objective is to design a guidance strategy for the nonlinear engagement dynamics outlined in \Cref{eq:1a1b1c}. The goal is to guarantee successful interception of the target at a predefined impact time, $t_d$, while ensuring that the following constraints are satisfied throughout the engagement:
\par{\textbf{Impact Time Convergence}:} The impact time error, $\varrho$, must converge to zero as $t \rightarrow t_d$, that is, $\varrho \rightarrow 0$ as $t \rightarrow t_d$.
\par{\textbf{Field-of-View (FOV) Constraint}:} The lead angle $\sigma$ must always remain within the bounds of the interceptor’s field of view, ensuring that $|\sigma| \leq \sigma_{\max}$ throughout the engagement.
\par{\textbf{Control Input Constraint}:} The lateral acceleration $a_M$ must not exceed the maximum allowable bound, satisfying the condition $|a_M| \leq a_{\max}$ at all times.  
\end{problem}
\begin{remark}
Note that the guidance strategy should be derived within a nonlinear framework to ensure its effectiveness across a wider range of initial heading angle errors.
\end{remark}
\section{Main Results}\label{sec3}
In this section, we propose a planar nonlinear guidance strategy designed to ensure target interception at a specified impact time while respecting the Field of View (FOV) constraints. Additionally, the strategy guarantees that the lateral acceleration remains within permissible bounds. To achieve this, we first present a method that ensures both continuous adherence to the FOV constraints throughout the engagement and interception at the desired impact time. Next, we introduce an input saturation model that bounds the lateral acceleration to prevent infeasible values.
Existing literature has extensively explored the use of Barrier Lyapunov functions in guidance strategy design \cite{rout2020modified}, as well as FOV-based impact time guidance \cite{kumar2021three}. These approaches provide a foundation for addressing the current problem.
In our design, we will utilize some existing results, which are summarized below.
\begin{lemma} (\citep{tee2009barrier}) \label{lem1}
 For any positive constants $k_{a_1}, k_{b_1}$,
 let $Z_1 := \left\{z_1 \in \mathbb{R}: -k_{a_1}<z_1 <k_{b_1}\right\}\subset \mathbb{R}$ and $\mathbb{N} := \mathbb{R}^l \times \mathbb{Z}_1 \subset \mathbb{R}^{l+1}$ be open sets. Consider the system $\dot \eta= h(t,\eta),$ where $\eta := [w,~~z_1]^T\in \mathbb{N}$, and $h : \mathbb{R}_+ \times \mathbb{N} \to \mathbb{R}^{l+1}$ is piece-wise continuous in $t$ and locally Lipschitz in $\eta$, uniformly in $t$, on $\mathbb{R}_+ \times \mathbb{N}$. Suppose that there exist functions $U : \mathbb{R}^l \to \mathbb{R}_+$ and $V_1: Z_1 \to \mathbb{R}_+$, continuously differentiable and positive definite in their respective domains, such that $V_1(z_1)\to\infty~\text{as}~ z_1\to -k_{a_1}~\text{or}~z_1\to k_{b_1},$ and $\gamma_1(||w||)\le U(w)\le \gamma_2(||w||),$ where $\gamma_1$ and $\gamma_2$ are class $K_{\infty}$ functions. Let $V(\eta) := V_1(z_1) + U(w),$ and $z_1(0)$ belong to the set $z_1 \in (-k_{a_1}, k_{b_1})$. If the inequality holds: $\dot V = -\dfrac{\partial V}{\partial \eta}h\le 0$, then $z_1(t)$ remains in the open set $z_1 \in (-k_{a_1}, k_{b_1})~\forall\,\,t \,\in [0,\infty)$.
\end{lemma}
\begin{lemma} (\cite{ren2010adaptive})\label{lem2}
For all $|\zeta|<1$ and any positive integer $p$, the following inequality holds  $\ln\left(\dfrac{1}{1-\zeta^{2p}}\right) <\dfrac{\zeta^{2p}}{1-\zeta^{2p}}$.
\end{lemma}
\subsection{Guidance Strategy with Field-of-View Constraints}\label{sec3A}
As discussed in \cite{assrkdm}, the time-to-go estimate for large heading angle errors in a two-dimensional setting is given by
\begin{equation}\label{t_go}
 t_{\rm go} = \frac {r}{V_M} \left( 1 + \frac{\sin^2\sigma}{\kappa} \right), \quad \kappa = 2(2\mathcal{N} - 1),
\end{equation}
where $ r $ is the radial distance, and $V_M$ is the interceptor's speed. This estimate applies to interceptors using proportional navigation, where the guidance law is given by $a_M = \mathcal{N} V_M \dot{\theta}_L$, with $\dot{\theta}_L$ being the Line-of-Sight (LOS) rate, and $\mathcal{N}$ is the navigation constant.

Define the time-to-go error as $\varrho := t_{\rm go} - t_{\rm go}^d$, where $t_{\rm go}^d = t_d - t_{\rm el}$ is the desired time-to-go, with $t_d$ being the impact time. From \Cref{t_go}, $t_{\rm go} = 0$ when $r = 0$, ensuring interception at the desired time for $\varrho = 0$.

Next, we define the upper bound on time-to-go as
$ t_{\rm go}^M = \frac{r}{V_M} \left( 1 + \frac{\sin^2\sigma_{\max}}{\kappa} \right)$
where $\sigma_{\max}$ is the maximum lead angle, and the minimum time-to-go is $t_{\rm go}^m = \frac{r}{V_M}$ for $\sigma = 0$, assuming unbounded lateral acceleration.

Let $\varrho_1 := t_{\rm go}^M - t_{\rm go}^d$ and its time derivative be given by
\begin{equation}\label{baredot}
 \dot{\varrho}_1 = 1 + \frac{\dot{r}}{V_M} \left( 1 + \frac{\sin^2\sigma_{\max}}{\kappa} \right).
\end{equation}
Using $\dfrac{\dot{r}}{V_M} = -\cos\sigma$, we get
 $\dot{\varrho}_1 = 1 - \cos\sigma - \dfrac{\cos\sigma \sin^2\sigma_{\max}}{\kappa}.$
Similarly, define $\varrho_2 := t_{\rm go}^m - t_{\rm go}^d$, with derivative given by 
$\dot{\varrho}_2 = 1 - \cos\sigma.$
For further details, refer to \cite{kumar2021three}.

\begin{remark}
If the impact time error satisfies \( \varrho_2 \leq \varrho \leq \varrho_1 \) for all \( t > 0 \), the interceptor's lead angle remains within some stipulated FOV bounds. Further, the bounds on time-to-go error depend dynamically on the current lead angle and its feasible range.
\end{remark}
\subsection{Guidance Strategy for Planar Engagements}\label{sec3B}
In this subsection, we develop the guidance strategy to address Problem \ref{prob}. We begin by deriving the dynamics of the lead angle using the expression for $\sigma$ in \Cref{eq:1a1b1c}, which yields
\begin{equation}\label{eq9}
\dot{\sigma} = \frac{a_M}{V_M} - \dot{\theta}_L = \frac{a_M}{V_M} + \frac{V_M \sin\sigma}{r}.
\end{equation}

Next, by differentiating $t_{\rm go}$ with respect to time, the time-to-go error's derivative turns out to be as follows:
\begin{equation}\label{eq:edot221}
\dot{\varrho} = \dot{t}_{\rm go} - \dot{t}_{\rm go}^d = 1 + \dot{t}_{\rm go} = 1 + \frac{\dot{r}}{V_M} \left( 1 + \frac{\sin^2\sigma}{\kappa} \right) + \frac{2r \sin\sigma \cos\sigma \dot{\sigma}}{\kappa V_M}.
\end{equation}

Substituting \Cref{eq9} into \Cref{eq:edot221}, we obtain
\begin{equation}
\dot{\varrho} = 1 - \cos\sigma \left( 1 + \frac{\sin^2\sigma}{\kappa} \right) + \frac{2 \sin^2\sigma \cos\sigma}{\kappa} + \frac{r \sin 2\sigma}{\kappa V_M^2} a_M = \mathcal{F}_p + \mathcal{G}_p a_M,
\end{equation}
where
\begin{align*}
\mathcal{F}_p &= 1 - \cos\sigma \left( 1 + \frac{\sin^2\sigma}{\kappa} \right) + \frac{2 \sin^2\sigma \cos\sigma}{\kappa} = 1 - \cos\sigma \left( 1 - \frac{\sin^2\sigma}{\kappa} \right)\\
 \mathcal{G}_p &= \frac{r \sin 2\sigma}{\kappa V_M^2}.
\end{align*}

To ensure the impact time error remains within bounds $\varrho_2 \leq \varrho \geq \varrho_1$, we may employ the barrier Lyapunov function-based backstepping technique, as outlined in \cite{kumar2021three}, or other alternative methods. Additionally, we aim to keep the interceptor's lateral acceleration within allowable bounds. To achieve this, we first introduce the input saturation model, which will be incorporated into the guidance design.

\subsubsection{Input Saturation Model}\label{sec3C}
In real-world applications, interceptors are often subject to actuator constraints, limiting available control inputs. To ensure the lateral acceleration remains within these bounds, we propose an input saturation model \cite{zou2019finite} given by
\begin{equation}\label{eq:ACC_a_dot}
    \dot{a}_M = \left[ 1 - \left( \frac{a_M}{a_{\max}} \right)^n \right] a_M^c - \rho a_M,
\end{equation}
where  $n \geq 2$ is even, and $\rho > 0$ is a small constant. In \Cref{eq:ACC_a_dot}, $a_M$ is the lateral acceleration control input, and $a_M^c$ is the commanded lateral acceleration. The term $-\rho a_M$ ensures that $\dot{a}_M$ approaches zero before $a_M$ reaches its maximum bound, $a_{\max}$. This introduces a safety margin between the maximum achievable acceleration and $a_{\max}$. To demonstrate that $a_M$ never reaches $a_{\max}$, we present the following theorem.
\begin{theorem}\label{InpThm}
Consider the input saturation model given in \Cref{eq:ACC_a_dot}. If the commanded input $a_M^c$ remains bounded for all $t \geq 0$, then the lateral acceleration $a_M$ will remain confined to the set
$\mathcal{S}_{a_M} := \{ a_M \colon |a_M| < a_{\max} \}$
for all $t \geq 0$.
\end{theorem}
\begin{proof}
The boundedness of the commanded input implies that there exists a constant $\mathcal{U}_M > 0$ such that $|a_M^c| \leq \mathcal{U}_M$ for all $t \geq 0$. Suppose, for the sake of contradiction, that $|a_M| = a_{\max}$. Then the input saturation model in \Cref{eq:ACC_a_dot} reduces to
$\dot{a}_M = -\rho a_M$.
Thus, $\dot{a}_M$ will be negative, causing $|a_M|$ to decrease. This shows that $|a_M| \leq a_{\max}$.
Next, we prove that $|a_M|$ will never reach \( a_{\max} \). Consider the input saturation model from \Cref{eq:ACC_a_dot}. Note that $1 - \left( \frac{a_M}{a_{\max}} \right)^n \geq 0$ for $|a_M| \in [0, a_{\max}]$. For $|a_M|$ to increase, both $a_M$ and $a_M^c$ must have the same sign, that is, $a_M a_M^c \geq 0$. This is because if $a_M$ and $a_M^c$ have opposite signs, then $\dot{a}_M>0$ for $a_M<0$ and $\dot{a}_M<0$ for $a_M>0$, causing $|a_M|$ to decrease. It may be readily checked from \Cref{eq:ACC_a_dot} that if $a_M > 0$ and $a_M^c < 0$, then $\dot{a}_M < 0$. Similarly, if $a_M < 0$ and $a_M^c > 0$, then $\dot{a}_M > 0$. Therefore, $|a_M|$ can only increase when $a_M$ and $a_M^c$ have the same sign.
Without loss of generality, assume that  $a_M^c > 0$ and $a_M \geq 0$. Since $a_M^c \leq \mathcal{U}_M$, we can use this to rewrite \Cref{eq:ACC_a_dot} as
\begin{equation}\label{eq:dotaMPrrof3}
\dot{a}_M \leq \left[ 1 - \left( \frac{a_M}{a_{\max}} \right)^n \right] \mathcal{U}_M - \rho a_M = \left[ 1 - \left( \frac{a_M}{a_{\max}} \right)^n - \frac{\rho a_{\max}}{\mathcal{U}_M} \frac{a_M}{a_{\max}} \right] \mathcal{U}_M.
\end{equation}
Using the inequality $ - \frac{a_M}{a_{\max}} \leq - \left( \frac{a_M}{a_{\max}} \right)^n $, we can further simplify this expression to the following:
\begin{equation}\label{eq:dotaMPrrof4}
\dot{a}_M \leq \mathcal{U}_M \left[ 1 - \left( 1 + \frac{\rho a_{\max}}{\mathcal{U}_M} \right) \left( \frac{a_M}{a_{\max}} \right)^n \right].
\end{equation}
From this, it follows that \( \dot{a}_M \leq 0 \) if we satisfy the condition
$\left[ 1 - \left( 1 + \frac{\rho a_{\max}}{\mathcal{U}_M} \right) \left( \frac{a_M}{a_{\max}} \right)^n \right] \leq 0$,
which further implies that
$a_M \geq a_{\max} \left( \frac{\mathcal{U}_M}{\mathcal{U}_M + \rho a_{\max}} \right)^{1/n}$.
Define \( \delta_M \) as
$\delta_M := a_{\max} \left( \frac{\mathcal{U}_M}{\mathcal{U}_M + \rho a_{\max}} \right)^{1/n}$,
where $ \delta_M < a_{\max}$. Substituting $a_M = \delta_M$ into the expression for $\dot{a}_M$, we obtain
 $ \dot{a}_M < 0$.
Thus, $|a_M| \leq \delta_M < a_{\max}$, which implies that $a_M$ never reaches the upper bound $a_{\max}$. Therefore, we conclude that $|a_M| < a_{\max}$ for all $t \geq 0$, and $a_M$ will never reach $a_{\max}$.
\end{proof}
\begin{figure}[!ht]
\begin{center}
    \includegraphics[width=\linewidth]{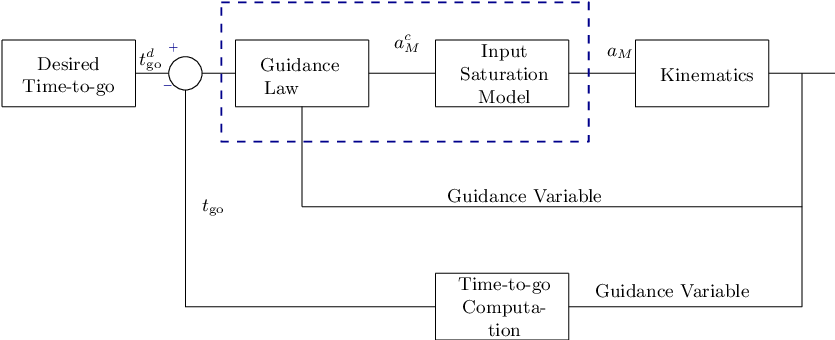}
\end{center}
    \caption{Schematic of FOV and input constrained impact time guidance strategy.}
\label{fig:block}
\end{figure}
\subsubsection{Guidance Approach with FOV and Input Constraint}\label{sec3D}
In the planar case, if the lateral acceleration, $a_M$, is treated as an additional state, the resulting system has the commanded input $a_M^c$. Thus, we can append \Cref{eq:ACC_a_dot} to the interceptor's kinematic equations and proceed to design the guidance law, which will be expressed as a function of $a_M^c$. The overall augmented system (as shown in \Cref{fig:block}) is as follows:
\begin{subequations}\label{eq:eq10}
\begin{align}
\dot{\varrho} &= \mathcal{F}_p + \mathcal{G}_p a_M, \label{eq:2ndOrderDynA}\\
\dot{a}_M &= \left[ 1 - \left( \frac{a_M}{a_{\max}} \right)^n \right] a_M^c - \rho a_M, \label{eq:2ndOrderDynB}
\end{align}
\end{subequations}
where $a_M^c$ is the commanded input. The augmented system in \Cref{eq:eq10} is a second-order dynamical system with a relative degree of two with respect to $a_M^c$. The objective is to design $a_M^c$ to ensure target interception at the desired impact time, while satisfying the field-of-view (FOV) constraints and respecting the bounds on lateral acceleration.
\begin{remark}
It is important to note that in our design, the guidance command (lateral acceleration) is obtained as the output of the input saturation model in \Cref{eq:2ndOrderDynB}, rather than directly designing $a_M$.
\end{remark}
\begin{theorem}\label{thm1}
Consider the interceptor-target engagement system is described by the dynamics in \Cref{eq:1a1b1c}. The commanded lateral acceleration, $a_M^c$, that guarantees target interception at a desired impact time, while adhering to field-of-view (FOV) constraints and respecting the bounds on lateral acceleration, is given by
\begin{equation}\label{a_Mcselect}
a_M^c = \frac{\rho a_M + \dot{\alpha}_1 - \mu \mathcal{G}_p \varrho^{2p-1} - \bar{\kappa}_3 \bar{z}_2}{\left[ 1 - \left(\frac{a_M}{a_{\max}} \right)^n \right]},
\end{equation}
where $\mu$ and $\alpha_1$ are defined as
\begin{equation}
\mu = \frac{q}{\varrho_1^{2p} - \varrho^{2p}} + \frac{(1 - q)}{\varrho_2^{2p} - \varrho^{2p}}, \quad 
\alpha_1 = \frac{1}{\mathcal{G}_p} \left[ - \mathcal{F}_p - \left( \bar{\kappa}_1 + \bar{\kappa}_2(t) \right) \varrho \right], \label{mu11Alpha1}
\end{equation}
where $\varrho_1 = t_{\rm go}^M - t_{\rm{go}}^{d}$, $\varrho_2 = t_{\rm go}^m - t_{\rm{go}}^{d}$, and $\dot{\alpha}_1$ is given by
\begin{align}
\nonumber \dot{\alpha}_1 &= \kappa V^2_M\dfrac{-\left[\sin\sigma- \dfrac{\sin^3\sigma}{\kappa} + \dfrac{\sin2\sigma\cos\sigma}{\kappa}\right]\dot\sigma -(\bar{\kappa}_1+\bar{\kappa}_2(t))\left[1 - \cos\sigma\left(1-\dfrac{\sin^2\sigma}{\kappa}\right)+ \dfrac{r\sin2\sigma}{\kappa V^2_M}a_M\right]}{r\sin2\sigma}\\
&+\dfrac{-\dot{\bar{\kappa}}_2 \varrho - (-V_M\cos\sigma\sin2\sigma +
2r\cos2\sigma\dot\sigma)\left(-1 + \cos\sigma - \cos\sigma\sin^2\sigma -r + \dfrac{\sin^2\sigma}{\kappa V_M} - t\right)\kappa V^2_M}{r\sin2\sigma}.
\end{align}
Additionally, the guidance law ensures that the impact time error $\varrho$ remains within the bounds given by
\begin{equation}\label{bounds}
\varrho_L(t) \leq \varrho(t) \leq \varrho_U(t), \quad \forall t > 0,
\end{equation}
where the lower and upper bounds on the impact time error are given by
\begin{equation}
\varrho_L(t) = \varrho_2 \left[ 1 - \exp \left( -2p \bar{\mathscr{V}}(0) \exp \left( -2p \bar{\kappa}_1 t \right) \right) \right],
\quad 
\varrho_U(t) = \varrho_1 \left[ 1 - \exp \left( -2p \bar{\mathscr{V}}(0) \exp \left( -2p \bar{\kappa}_1 t \right) \right) \right].
\end{equation}
Moreover, the impact time error $\varrho$ converges to zero at the time of interception, provided that the control gains satisfy the conditions
$\bar{\kappa}_1 > 0, \quad \rho > 0, \quad \bar{\kappa}_3 > 0$,
while the gain $\bar{\kappa}_2$ satisfies
\begin{equation}\label{k2}
\bar{\kappa}_2 = \sqrt{\bar{\beta} + \left( \frac{\dot{\varrho}_1}{\varrho_1} \right)^2 + \left( \frac{\dot{\varrho}_2}{\varrho_2} \right)^2} > 0, \quad \bar{\beta} > 0.
\end{equation}
Therefore, the interceptor will achieve target interception at the desired impact time while satisfying both the field-of-view and lateral acceleration constraints.
\end{theorem}

\begin{proof}
We begin by considering the asymmetric, time-varying barrier Lyapunov function proposed in \cite{tee2009barrier, tee2011control}:

\begin{equation}\label{v}
\mathscr{V}_1 = \frac{q(\varrho)}{2} \ln \left( \frac{\varrho_1^{2p}}{\varrho_1^{2p} - \varrho^{2p}} \right) + \frac{(1 - q(\varrho))}{2} \ln \left( \frac{\varrho_2^{2p}}{\varrho_2^{2p} - \varrho^{2p}} \right),
\end{equation}
where $p \geq 1$ is an integer, and $ q(\varrho)$ is defined as
$q(\varrho) = \begin{cases}
1, & \varrho > 0, \\
0, & \varrho \leq 0.
\end{cases}$
Next, we introduce the terms $\varrho_a$ and $\varrho_b$, defined as
$\varrho_b = \frac{\varrho}{\varrho_1}$, $\varrho_a = \frac{\varrho}{\varrho_2}$,
and the convex combination
 $\bar{\varrho} = q \varrho_b + (1 - q) \varrho_a$.
We also define auxiliary states $\bar{z}_1$ and $\bar{z}_2$ as
$\bar{z}_1 = \varrho = t_{\rm go} - t^d_{\rm go}, \quad \bar{z}_2 = a_M - \alpha_1$,
with the time derivative of $\bar{z}_2$ related as
$\dot{\bar{z}}_2 = \dot{a}_M - \dot{\alpha}_1$.
The Lyapunov function $\mathscr{V}_1$ may now be rewritten as
\begin{equation*}
\mathscr{V}_1 = \frac{q}{2p} \ln \left( \frac{1}{1 - \frac{\varrho^{2p}}{\varrho_1^{2p}}} \right) + \frac{(1 - q)}{2p} \ln \left( \frac{1}{1 - \frac{\varrho^{2p}}{\varrho_2^{2p}}} \right)
= \frac{q}{2p}\ln \left[\frac{1}{1-\varrho^{2p}_b} \right]+ \frac{1-q}{2p}\ln\left[ \frac{1}{1-\varrho^{2p}_a}\right],
\end{equation*}
which simplifies to
\begin{equation}\label{eqReLyap}
\mathscr{V}_1 = - \frac{q}{2p} \ln (1 - \varrho_b^{2p}) - \frac{1 - q}{2p} \ln (1 - \varrho_a^{2p})\\
\end{equation}
This expression can be compactly written in terms of $\bar{\varrho}$ as $\mathscr{V}_1 = \frac{1}{2p} \ln \left( \frac{1}{1 - \bar{\varrho}^{2p}} \right)$
which is positive definite and continuously differentiable within the region $|\varrho_1| \leq 1$ \cite{tee2009barrier, tee2011control}.
Next, we differentiate $ \mathscr{V}_1$ in \Cref{eqReLyap} with respect to time to obtain  $\dot{\mathscr{V}}_1$ as
\begin{align}\label{eq12}
\nonumber\dot{\mathscr{V}}_1 &= \left(\frac{q}{2p}\frac{2p\varrho^{2p-1}_b}{1-\varrho^{2p}}\right)\dot{\varrho}_b +  \left[\frac{(1-q)}{2p}\frac{2p\varrho^{2p-1}_a}{1-\varrho^{2p}_a}\right]\dot{\varrho}_a = \nonumber  \frac{q\varrho^{2p-1}_b}{1-\varrho^{2p}_b}\left[ \frac{\dot{\varrho}}{\varrho_1}-\frac{\varrho\dot{\varrho}_1}{\varrho_1^2}\right ] +  \frac{(1-q)\varrho^{2p-1}_a}{1-\varrho^{2p}_a}\left[ \frac{\dot{\varrho}}{\varrho_2}-\frac{\varrho\dot{\varrho}_2}{\varrho_2^2}\right ],\\
&= \dfrac{q \varrho_b^{2p - 1}}{\varrho_1 (1 - \varrho_b^{2p})} \left( \dot{\varrho} - \dfrac{\varrho \dot{\varrho}_1}{\varrho_1} \right) + \frac{(1 - q) \varrho_a^{2p - 1}}{\varrho_2 (1 - \varrho_a^{2p})} \left( \dot{\varrho} - \frac{\varrho \dot{\varrho}_2}{\varrho_2} \right),
\end{align}
whereupon the dynamics of the system, given by \Cref{eq:2ndOrderDynA} and \Cref{eq:2ndOrderDynB}, may be substituted into the above expression to result in the following expression for $\dot{\mathscr{V}}_1$:
\begin{equation}\label{eq13}
\dot{\mathscr{V}}_1 = \frac{q \varrho_b^{2p - 1}}{\varrho_1 (1 - \varrho_b^{2p})} \left( \mathcal{F}_p + \mathcal{G}_p (\bar{z}_2 + \alpha_1) - \varrho \frac{\dot{\varrho}_1}{\varrho_1} \right) + \frac{(1 - q) \varrho_a^{2p - 1}}{\varrho_2 (1 - \varrho_a^{2p})} \left( \mathcal{F}_p + \mathcal{G}_p (\bar{z}_2 + \alpha_1) - \varrho \frac{\dot{\varrho}_2}{\varrho_2} \right).
\end{equation}
Substituting $ \alpha_1 $ from \Cref{mu11Alpha1} and $ \bar{\kappa}_2 $ from \Cref{k2} into the expression above, one may arrive at
\begin{equation} \label{putEaEb}
\dot{\mathscr{V}}_1 = \frac{q \varrho_b^{2p - 1}}{\varrho_1 (1 - \varrho_b^{2p})} \left( \mathcal{G}_p \bar{z}_2 - \bar{\kappa}_1 \varrho - \bar{\kappa}_2(t) \varrho - \varrho \frac{\dot{\varrho}_1}{\varrho_1} \right)
+ \frac{(1 - q) \varrho_a^{2p - 1}}{\varrho_2 (1 - \varrho_a^{2p})} \left( \mathcal{G}_p \bar{z}_2 - \bar{\kappa}_1 \varrho - \bar{\kappa}_2(t) \varrho - \varrho \frac{\dot{\varrho}_2}{\varrho_2} \right).
\end{equation}
Upon further simplification, substituting $ \varrho_b = \left({\varrho}/{\varrho_1}\right) $ and $ \varrho_a = \left({\varrho}/{\varrho_2}\right) $, we obtain
\begin{align}\nonumber
\dot{\mathscr{V}}_1 &= -\bar{\kappa}_1 \left( \frac{q \varrho_b^{2p}}{1 - \varrho_b^{2p}} + \frac{(1 - q) \varrho_a^{2p}}{1 - \varrho_a^{2p}} \right)
+ \mathcal{G}_p \bar{z}_2 \left( \frac{q \varrho_b^{2p}}{\varrho (1 - \varrho_b^{2p})} + \frac{(1 - q) \varrho_a^{2p}}{\varrho (1 - \varrho_a^{2p})} \right)\\
&+ \bar{\kappa}_2(t) \left( - \frac{q \varrho_b^{2p}}{1 - \varrho_b^{2p}} - \frac{(1 - q) \varrho_a^{2p}}{1 - \varrho_a^{2p}} \right)
- \frac{\dot{\varrho}_1}{\varrho_1}\frac{q\varrho^{2p}_b}{(1-\varrho^{2p}_b)} - \frac{\dot{\varrho}_2}{\varrho_2}\frac{(1-q)\varrho^{2p}_a}{(1-\varrho^{2p}_a) }.
\end{align}
Imposing the relation $ \bar{\varrho} = q\varrho_b + (1-q)\varrho_a $, we may express $\dot{\mathscr{V}}_1$ as follows:
\begin{equation*}
\dot{\mathscr{V}}_1 = \frac{-\bar{\kappa}_1\bar{\varrho}^{2p}}{1-\bar{\varrho}^{2p}} + \mu_1\mathcal{G}_p\varrho^{2p-1}\bar{z}_2 
    - \left[\bar{\kappa}_2(t)\frac{\bar{\varrho}^{2p}}{1-\bar{\varrho}^{2p}}+\frac{\dot{\varrho}_1}{\varrho_1}\frac{q\varrho^{2p}_b}{(1-\varrho^{2p}_b)}+\frac{\dot{\varrho}_2}{\varrho_2}\frac{(1-q)\varrho^{2p}_a}{(1-\varrho^{2p}_a) }\right].
\end{equation*}
By choosing $\bar{\kappa}_2(t)$ as specified in \Cref{k2} and noting that $q \in \{0,1\}$, we conclude that 
$\bar{\kappa}_2 + q\left({\dot{\varrho}_1}/{\varrho_1}\right) + (1-q)\left({\dot{\varrho}_2}/{\varrho_2}\right) \geq 0$.
Thus, the expression for \( \dot{\mathscr{V}}_1 \) simplifies to 
$\dot{\mathscr{V}}_1 \leq -\left({\bar{\kappa}_1\bar{\varrho}^{2p}}/({1-\bar{\varrho}^{2p}})\right) + \mu_1\mathcal{G}_p\varrho^{2p-1}\bar{z}_2$,
where $ \mu_1$ is defined in \Cref{mu11Alpha1}. 

To enforce the constraints on the interceptor's lateral acceleration, we augment $\mathscr{V}_1$ with a quadratic Lyapunov candidate $\mathscr{V}_2$. The overall Lyapunov function candidate for the augmented system, as given in \Cref{eq:eq10}, is therefore
$\bar{\mathscr{V}} = \mathscr{V}_1 + \mathscr{V}_2$,
where $\mathscr{V}_2 = 0.5\bar{z}_2^2$. Differentiating $\bar{\mathscr{V}}$ and substituting for $\dot{\mathscr{V}}_1$, we obtain
\begin{align}
    \nonumber \dot{\bar{\mathscr{V}}} &= \dot{\mathscr{V}}_1 + \bar{z}_2 \dot{\bar{z}}_2 \\
    &\le \mu_1 \mathcal{G}_p \varrho^{2p-1} \bar{z}_2 - \bar{\kappa}_1 \left( \frac{\bar{\varrho}^{2p}}{1-\bar{\varrho}^{2p}} \right) + \bar{z}_2 \left\{ \left[ 1 - \left( \frac{a_M}{a_{\max}} \right)^n \right] a_M^c - \rho a_M - \dot{\alpha}_1 \right\}, \\
    &\le - \bar{\kappa}_1 \left( \frac{\bar{\varrho}^{2p}}{1-\bar{\varrho}^{2p}} \right) + \bar{z}_2 \left\{ \left[ 1 - \left( \frac{a_M}{a_{\max}} \right)^n \right] a_M^c - \rho a_M - \dot{\alpha}_1 + \mu_1 \mathcal{G}_p \varrho^{2p-1} \right\}.
    \label{eq:T7_LyaFuncDot_31}
\end{align}
Substituting the expression for the commanded acceleration $a_M^c$ from \Cref{a_Mcselect} into \Cref{eq:T7_LyaFuncDot_31}, one may get
$$\dot{\bar{\mathscr{V}}} \leq -\bar{\kappa}_1 \left( \frac{\bar{\varrho}^{2p}}{1 - \bar{\varrho}^{2p}} \right) - \bar{\kappa}_3 \bar{z}_2^2,
$$ 
where $\bar{\kappa}_3 > 0$ is a constant. By utilizing the results from \Cref{lem2}, we may express  $\dot{\bar{\mathscr{V}}}$ in a more compact form as
\begin{equation} \label{eq:T7_LyaFuncDot_33}
    \dot{\bar{\mathscr{V}}} \leq -\bar{\kappa}_1 \ln \left( \frac{1}{1 - \bar{\varrho}^{2p}} \right) - \bar{\kappa}_3 \bar{z}_2^2.
\end{equation}
From the definitions of $\mathscr{V}_1$ and $\mathscr{V}_2$, \Cref{eq:T7_LyaFuncDot_33} simplifies to $
\dot{\bar{\mathscr{V}}} \leq -2p \bar{\kappa}_1 \mathscr{V}_1 - 2 \bar{\kappa}_3 \mathscr{V}_2 \leq -2 \kappa_p \bar{\mathscr{V}},
$ where $\kappa_p = \min \{ p\bar{\kappa}_1, \bar{\kappa}_3 \} $.
Integrating \( \dot{\bar{\mathscr{V}}} \) over appropriate limits, we arrive at 
$\bar{\mathscr{V}} \leq \bar{\mathscr{V}}(0) \exp \left( -2 \kappa_p t \right)$, $ \forall t > 0$,
which may be further rewritten as $$\frac{1}{2p} \left( \ln \left( \frac{1}{1 - \bar{\varrho}^{2p}} \right) + 2p \cdot \frac{1}{2} \bar{z}_2^2 \right) \leq \bar{\mathscr{V}}(0) \exp \left( -2 \kappa_p t \right).$$
Since \( \bar{\varrho} \leq 1 \) and \( \bar{z}_2^2 \geq 0 \), we may write
\begin{subequations} \label{eq:fConstr}
 \begin{align}
    \frac{1}{2p} \ln \left( \frac{1}{1 - \bar{\varrho}^{2p}} \right) \leq \bar{\mathscr{V}}(0) \exp \left( -2 \kappa_p t \right) &\implies \bar{\varrho}^{2p} \leq 1 - \exp \left[ -2p \bar{\mathscr{V}}(0) \exp \left( -2 \kappa_p t \right) \right], \label{eq:fConstr1} \\
    \frac{1}{2} \bar{z}_2^2 \leq \bar{\mathscr{V}}(0) \exp \left( -2 \kappa_p t \right) &\implies \bar{z}_2 \leq \pm \sqrt{2 \bar{\mathscr{V}}(0) \exp \left( -2 \kappa_p t \right)}. \label{eq:fConstr2}
\end{align}
\end{subequations}
Using \Cref{eq:fConstr1}, it may now be shown that $\varrho_L(t) \leq \varrho(t) \leq \varrho_U(t) \quad \forall t > 0,$ where $\varrho_L(t)$ and $\varrho_U(t)$ are defined in \Cref{bounds}. 
Given the choice of gains $\bar{\kappa}_1$ and $\bar{\kappa}_3$, and with $p = 1$, it follows that the right-hand side of \Cref{eq:T7_LyaFuncDot_33} is negative. Furthermore, substituting \Cref{eq:fConstr2} into the relation $\bar{z}_2 = a_M - \alpha_1$, we obtain 
$\alpha_1 = a_M \mp \sqrt{2 \bar{\mathscr{V}}(0) \exp \left( -2 \kappa_p t \right)}$.

Since both $a_M$ (as given in \Cref{InpThm}) and $\sqrt{2 \bar{\mathscr{V}}(0) \exp \left( -2 \kappa_p t \right)}$ are bounded, it follows that $\alpha_1$ will also remain bounded. Thus, for a properly chosen gain, we may conclude that $\dot{\bar{\mathscr{V}}}$ is negative definite. Consequently, the system is stable, target interception will be achieved at the desired impact time, and the state variables will stay within their respective bounds, converging to zero at the time of interception. The field of view (FOV) and the interceptor's lateral acceleration will also stay within their boundaries, ensuring all constraints are satisfied.
This completes the proof.
\end{proof}
\begin{theorem}
 The heading angle error $\sigma(t)$ converges to
zero in the vicinity of the target interception. 
\end{theorem}
\begin{proof}
Let us select a Lyapunov function candidate $\mathcal{V}_{\sigma} = 0.5\sigma^2$. Differentiating $\mathcal{V}_{\sigma}$, one may obtain $\dot{\mathcal{V}}_{\sigma} = \sigma\dot{\sigma}$. Next, since $\dot{\bar{\mathcal{V}}} \leq 0$ (from \Cref{eq:T7_LyaFuncDot_33}), it follows that $\dot{\varrho}$ is negative and $\varrho \rightarrow 0$. As a result, \Cref{eq:edot221} may be set to zero, yielding
\begin{equation}\label{eq:EdotSigma}
    \dot{\sigma} = -\dfrac{V_M\kappa}{r\sin{2\sigma}}\left[1 + \dfrac{\dot{r}}{V_M}\left(1+\dfrac{\sin^2}{\kappa}\right)\right].
\end{equation}
Substituting \Cref{eq:EdotSigma} in the expression of $\dot{\mathcal{V}}_{\sigma}$, one may obtain 
\begin{align}
    \dot{\mathcal{V}}_{\sigma} &= -\sigma\left\{\dfrac{V_M\kappa}{r\sin{2\sigma}}\left[1 + \dfrac{\dot{r}}{V_M}\left(1+\dfrac{\sin^2}{\kappa}\right)\right]\right\},\\
    &= -\frac{V_M}{r}\left(\frac{\sigma}{2}\sin{\left(\frac{\sigma}{2}\right)}\right)\left(\frac{\kappa}{\cos{\left(\frac{\sigma}{2}\right)} \cos{\sigma}} - 2\cos{\frac{\sigma}{2}}\right).
\end{align}
Note that since $V_M > 0$, $r > 0$, and $\kappa > 2$, along with the conditions 
$\dfrac{\sigma}{2}\sin\left(\dfrac{\sigma}{2}\right) > 0$ and $ \left(\frac{\kappa}{\cos{\left(\frac{\sigma}{2}\right)} \cos{\sigma}} - 2\cos{\frac{\sigma}{2}}\right) > 0$ holding for $|\sigma| \leq \pi/2$. Therefore, $\dot{\mathcal{V}}_{\sigma} < 0$, for $|\sigma| \leq \pi/2$, and the lead angle converge to zero near asymptotically. This concludes the proof.
\end{proof}
\begin{corollary}
The magnitude of the lateral acceleration reduces in accordance with the
lead angle, and goes to zero towards the end of the engagement, that is, near the target
interception.
\end{corollary}
\begin{proof}
The stabilizing function, when the lead angle $\sigma\rightarrow 0$, can be expressed as $\lim_{\sigma\to 0} \alpha_1$. From \Cref{eq:T7_LyaFuncDot_33}, as $\dot{\bar{\mathcal{V}}}$ is negative, it follows that  $\varrho\rightarrow 0$, hence 
\begin{align}
      \lim_{\sigma\to 0} \alpha_1  = -\lim_{\sigma\to 0}  \dfrac{\mathcal{F}_p}{\mathcal{G}_p} = - \lim_{\sigma\to 0} \dfrac{1 - \cos{\sigma}\left(1 + \dfrac{\sin^2{\sigma}}{\kappa}\right) + \dfrac{2\sin^2{\sigma}\cos{\sigma}}{\kappa}}{r\sin{2\sigma}} \kappa V_{M}^{2}. 
    \end{align}
Furthermore, interceptors are typically equipped with warheads that detonate when $r < r_{\rm{lethal}}$, where $r_{\rm{lethal}}$ is the lethal radius (dependent on the warhead).
Consequently, the scenario of $r = 0$ may be ruled out. Therefore,
\begin{equation}\label{eq:limitAlpha}
\lim_{\sigma\to 0} \alpha_1  = -\lim_{\sigma\to 0} \dfrac{1 - \cos{\sigma}\left(1 - \dfrac{\sin^2{\sigma}}{\kappa}\right)}{r\sin{2\sigma}} \kappa V_{M}^{2} = -\lim_{\sigma\to 0} \dfrac{1 - \cos{\sigma} - \cos{\sigma}\dfrac{\sin^2{\sigma}}{\kappa}}{r\sin{2\sigma}} \kappa V_{M}^{2}.
\end{equation}
Since both the numerator and denominator of \Cref{eq:limitAlpha} tend to zero, therefore
\begin{equation*}
    \lim_{\sigma\to 0} \alpha_1 = -\lim_{\sigma\to 0} \dfrac{\sin{\sigma} + \sin{\sigma}\dfrac{\sin^2{\sigma}}{\kappa} - \cos{\sigma}\sin{2\sigma}}{2r\cos{2\sigma}} \kappa V_{M}^{2} = 0.
\end{equation*}
and hence $\dot{\alpha}_1 = 0$ as $\sigma\rightarrow 0$. Next, one may substitute the expression of $a_{M}^{c}$ in \Cref{eq:2ndOrderDynB} to get
\begin{equation*}
    \dot{a}_{M} = \dot{\alpha}_1 - \mu \mathcal{G}_p \varrho^{2p-1} - \bar{\kappa}_3 \bar{z}_2 = - \mu \mathcal{G}_p \varrho^{2p-1} - \bar{\kappa}_3 \bar{z}_2. 
\end{equation*}
From \Cref{eq:T7_LyaFuncDot_33}, it follows that $\bar{z}_2 \rightarrow 0$, hence, $\dot{a}_{M} =  - \mu \mathcal{G}_p \varrho^{2p-1}$, which in turn implies that $a_M\rightarrow 0$ near target interception. This completes the proof.
\end{proof}
\subsubsection{Extension of Proposed Strategy to a Large Impact Time in the presence of Input Constraints} \label{sec3E}
The guidance strategies derived in the previous subsections perform effectively as long as the desired impact time ensures that the time-to-go error remains within acceptable bounds, specifically $t_d \in [r/V_M, ~t_{\rm go}^M]$. However, in practice, there may be scenarios where the desired impact time exceeds the upper bound, $t_{\rm go}^M$, requiring an alternative approach. In such cases, a multi-stage guidance strategy can be employed, where additional stages are incorporated to allow the interceptor to take a detour, while ensuring that the target remains within the interceptor's field of view (FOV). It is important to note that there is an upper bound to the achievable impact time under the FOV constraints, given by $ \frac{r(0)}{V_M \cos\sigma_{\max}} $.
To intercept the target at a larger impact time, one approach is to initially implement the deviated pursuit strategy, as discussed in \cite{kumar2021three}, and then switch to the method outlined in \Cref{sec3D} at a specific time instant, which can be determined from the desired impact time $t_d$. Alternatively, the problem can be solved by increasing the heading error during the first stage, followed by maintaining a deviated pursuit course with a maximum lead angle, within the allowable FOV bound. The guidance strategy from \Cref{sec3D} is then used after the switch.

In the deviated pursuit strategy with lead angle $\sigma$, the relative separation between the interceptor and the target decreases at a constant rate of $V_M \cos\sigma$. Consequently, the relative separation will be reduced by $V_M t_1 \cos\sigma$ up to the switching time instant $t_1$. The switching instant $t_1$ can be derived as
\begin{align}\label{t1}
t_1 = \frac{V_M(t_d + \epsilon) - \Lambda r(0)}{V_M \left[1 - \Lambda \cos\sigma(0)\right]},
\end{align}
where $\epsilon$ is a small positive constant, and $\Lambda$ is a constant related to the pursuit strategy. For a given set of initial engagement conditions, interception at the desired impact time $t_d$ can be achieved if it satisfies the condition $ 0 \leq t_1 \leq t_d $.

As discussed in \cite{kumar2021three}, to achieve interception at a larger impact time, it is essential to maintain $ \dot{\sigma} = 0 $, meaning that the guidance command for the first stage should enforce a constant heading angle. However, this may bound the interceptor's lead angle to its initial value until the switching instant $t_1$, even if a larger permissible lead angle is available. Therefore, unlike in \cite{kumar2021three}, the guidance design in this work provides a desired lead angle value $\sigma_d$, such that $ \sigma(0) \leq \sigma_d < \sigma_{\max} $.
For a planar engagement scenario, as shown in \Cref{fig:1}, the guidance command for the deviated pursuit can be expressed as $a_M = V_M \dot{\theta}_L$ for the pitch plane. In the following theorem, we present the commanded lateral acceleration that ensures the interceptor follows the deviated pursuit course up to time $t_1$, while maintaining a bounded control input.

\begin{theorem}
Consider an interceptor-target engagement scenario as depicted in \Cref{fig:1}, with a desired lead angle $\sigma_d$ and a switching instant $t_1$ given by \Cref{t1}. For a desired impact time $t_d$ greater than the upper bound on the time-to-go estimate, that is, $t_d > t_{\rm go}^M$, the interceptor will achieve the desired lead angle $\sigma_d$ if its lateral acceleration command is designed as follows:
\begin{equation} \label{eq:A_MCThreeStage}
a_M^c = \frac{1}{\left[1 - \left(\frac{a_M}{a_{\max}}\right)^n\right]}\left(\rho a_M + V_M \ddot{\theta}_L - V_M \xi \frac{q_f}{p_f} \dot{\sigma}_e^{2 - \frac{p_f}{q_f}} - c \sign(s)\right),
\end{equation}
where $\xi \in \mathbb{R}_{+}$ is a positive real number, $p_f$ and $q_f$ are odd positive integers such that $p_f > q_f$ and $1 < \frac{p_f}{q_f} < 2$, $c > 0$ is a positive constant, and $\sign(s)$ denotes the sign function.
With this control law, the desired lead angle, $\sigma_d$, will be achieved within a finite time $t_s$, where $t_s < t_1$, and the control input remains bounded.
\end{theorem}
\begin{proof}
Let the error between the actual and desired lead angles be defined as $\sigma_e = \sigma - \sigma_d$. Differentiating $\sigma_e$ with respect to time gives $\dot{\sigma}_e = \dot{\sigma}$. Substituting $\dot{\sigma}$ from \Cref{eq:1a1b1c} into this, we obtain:
$\dot{\sigma}_e = \frac{a_M}{V_M} - \dot{\theta}_L$.
Similarly, the time derivative of $\dot{\sigma}_e$ is
$\ddot{\sigma}_e = \frac{\dot{a}_M}{V_M} - \ddot{\theta}_L$.
From \Cref{eq:eq10}, we have the expression for $\dot{a}_M$, and $\ddot{\theta}_L$ can be written as
$\ddot{\theta}_L = -\frac{2 \dot{r} \dot{\theta}_L}{r} - \frac{\cos \sigma}{r} a_M$.
Substituting this into the expression for $\ddot{\sigma}_e$, one may obtain
\begin{equation*}
    \ddot{\sigma}_e = \dfrac{\left[ 1-\left(\dfrac{a_M}{a_{\max}}\right)^n \right] a_{M}^c - \rho a_M}{V_M} - \ddot{\theta}_L= - \dfrac{2 \dot r\dot{\theta}_L}{r} -\left( \dfrac{\cos \sigma}{r} - \frac{\rho}{V_M}\right)a_M +\frac{1}{V_M}\left[1-\left(\dfrac{a_M}{a_{\max}}\right)^n \right]a_{M}^c.
\end{equation*}
Next, consider a sliding surface defined as
$s = \sigma_e + \frac{1}{\xi} \dot{\sigma}_e^{p_f/q_f}$,
where $\xi \in \mathbb{R}_{+}$, and $p_f$ and $q_f$ are odd positive integers with $p_f > q_f$ and $1 < \frac{p_f}{q_f} < 2$. This choice ensures that the term $\dot{\sigma}_e^{p_f/q_f}$ has a real root. The sliding surface is selected such that when the sliding mode is enforced on $s = 0$, both $\sigma_e$ and $\dot{\sigma}_e$ converge to zero, guaranteeing that the lead angle $\sigma$ reaches its desired value and remains there until the switching instant $t_1$ as given in \Cref{t1}.
We define a Lyapunov candidate function as $\mathscr{V} = ({s^2}/{2})$, and differentiate it to obtain $\dot{\mathscr{V}} = s \dot{s}$.
Differentiating $s$ with respect to time, we have
$\dot{s} = \dot{\sigma}_e + \frac{p_f}{\xi q_f} \dot{\sigma}_e^{\frac{p_f}{q_f} - 1} \ddot{\sigma}_e$.
Substituting this expression into $\dot{\mathscr{V}}$, one may arrive at
$\dot{\mathscr{V}} = s \left( \dot{\sigma}_e + \frac{p_f}{\xi q_f} \dot{\sigma}_e^{\frac{p_f}{q_f} - 1} \ddot{\sigma}_e \right).$
Next, substituting the expression for $\ddot{\sigma}_e$ in $\dot{\mathscr{V}}$ yields
\begin{equation}\label{eq:VdotThreeStage}
     \dot{\mathscr{V}} = \frac{s}{V_M}\left(\frac{1}{\xi}\frac{p_f}{q_f}\dot{\sigma}_e^{(p_f/q_f)-1}\left\{\left[1 - \left(\frac{a_M}{a_{\max}}\right)^n\right]a_M^c - \rho a_M - V_M\ddot{\theta}_L\right\} + V_M \dot{\sigma}_{e}\right)
\end{equation}
Choosing the commanded lateral acceleration $a_M^c$ as in \Cref{eq:A_MCThreeStage} and substituting it into \Cref{eq:VdotThreeStage} results in
\begin{equation}
 \dot{\mathscr{V}} = - \frac{s}{V_M} \frac{1}{\xi}\frac{p_f}{q_f}\dot{\sigma}_e^{(p_f/q_f)-1} c \sign(s)
  =  -\frac{1}{V_M\xi}\frac{p_f}{q_f}\dot{\sigma}_e^{(p_f/q_f)-1} c |s| = -K(\dot{\sigma}_e) |s|,
\end{equation}
where $K(\dot{\sigma}_e) = \frac{c p_f \dot{\sigma}_e^{\frac{p_f}{q_f} - 1}}{V_M q_f \xi}$, which is positive for all $s \neq 0$, since $p_f$ and $q_f$ are odd positive integers, and $\dot{\sigma}_e^{\frac{p_f}{q_f} - 1} > 0$ for all $\sigma_e \neq 0$. Thus, $\dot{\mathscr{V}}$ is negative for all $s \neq 0$, and the sliding mode is enforced on $s = 0$ within a finite time, say $t_r$.
Once the sliding mode is enforced, the reduced-order dynamics are given by $\sigma_e = -\frac{1}{\xi} \dot{\sigma}_e^{\frac{p_f}{q_f}}$,
which leads to $- \frac{\dot{\sigma}_e}{\sigma_e^{q_f/p_f}} = \xi$.
Solving for the time instant $t_s$ such that $\sigma_e = 0$ for all $t \geq t_s$, we obtain:
\begin{equation*}
\xi (t_s-t_r) = \int_{\sigma_e(t_r)}^0 \sigma_e^{-q_f/p_f} \dot{\sigma}_e\mathbf{d}t \implies t_s = t_r + \dfrac{p_f}{\xi(p_f-q_f)} \lvert \sigma_e(t_r) \rvert ^{1-q_f/p_f}.
\end{equation*}
This shows that the error variables $\sigma_e$ and $\dot{\sigma}_e$ will converge to zero within a finite time $t_s$, thereby completing the proof.
\end{proof}
\begin{remark}
Note that substituting \Cref{eq:A_MCThreeStage} into the input saturation model in \Cref{eq:ACC_a_dot} yields $\dot{a}_M = V_M\ddot{\theta}_L - (V_M \xi {q_f}/{p_f})\dot{\sigma}_e^{2 - (p_f/q_f)} - c \sign(s)$. Once the sliding mode is enforced, the terms $(V_M \xi {q_f}/{p_f})\dot{\sigma}_e^{2 - (p_f/q_f)}$ and $c \sign(s)$ vanish, leaving $\dot{a}_M = V_M\ddot{\theta}_L$. Integrating this within feasible bounds results in $a_M = V_M\dot{\theta}_L$.
\end{remark}
\section{Simulation Studies}\label{sec4}
In this section, we evaluate the performance of the proposed guidance strategy through several simulation scenarios. The interceptor is assumed to travel at a constant speed of $250 \text{m/s}$ and is subject to a maximum available lateral acceleration of $\pm a_{\max}$. The inertial reference frame is initially centered at the interceptor's center of gravity. The target is positioned at an initial radial distance of $10 \, \text{km}$, with a line-of-sight (LOS) angle, $\theta_L = 0^\circ$, relative to the interceptor.
For visualization purposes, we denote the target’s position with a star marker on the trajectory plots, and the interceptor’s initial location is indicated by a square marker.
The design parameters used for the simulations are as follows: $\bar{\kappa}_1 =1$, $\bar{\beta}=1$, $\mathcal{N}=3$, $\bar{\kappa}_3 =1$, $a_{\max} =\pm{20}g$, $\rho = 0.1$, $p_f = 11$, $q_f = 9$, $c=1000$, and $p=1$, while the gain $\bar{\kappa}_2$ is chosen in accordance with \Cref{k2}.
\subsection{Fixed impact time for various interceptor's heading angles error}
\begin{figure}[!ht]
\centering
\begin{subfigure}{0.33\linewidth}
	\centering
\includegraphics[width=\linewidth]{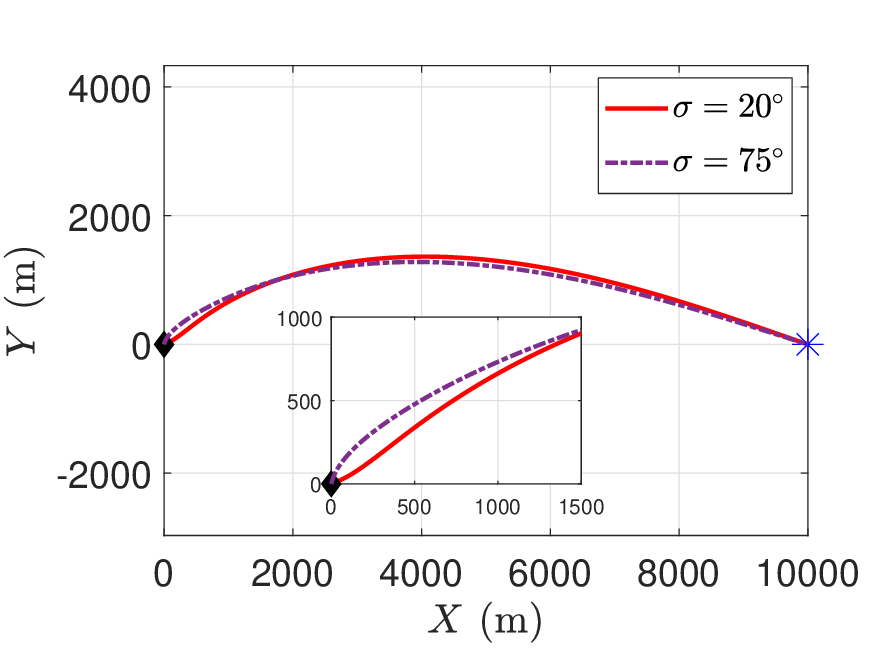}
    \caption{Trajectory.}
\label{fig:Time42DifferentSigma_XmYm}
\end{subfigure}%
\begin{subfigure}{0.33\linewidth}
	\centering
\includegraphics[width=\linewidth]{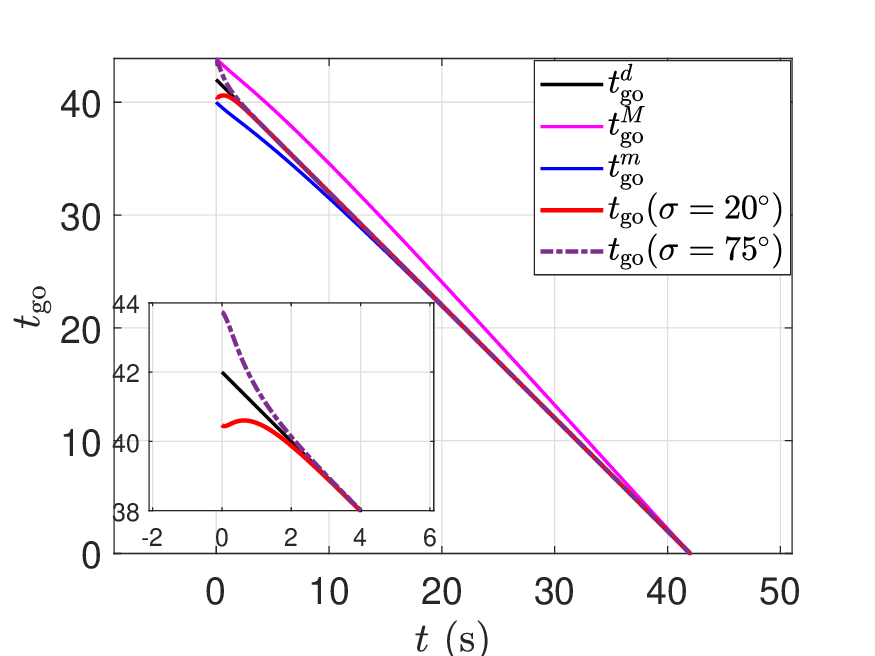}
    \caption{Time-to-go estimates.}
\label{fig:Time42DifferentSigma_Time}
\end{subfigure}%
\begin{subfigure}{0.33\linewidth}
	\centering
\includegraphics[width=\linewidth]{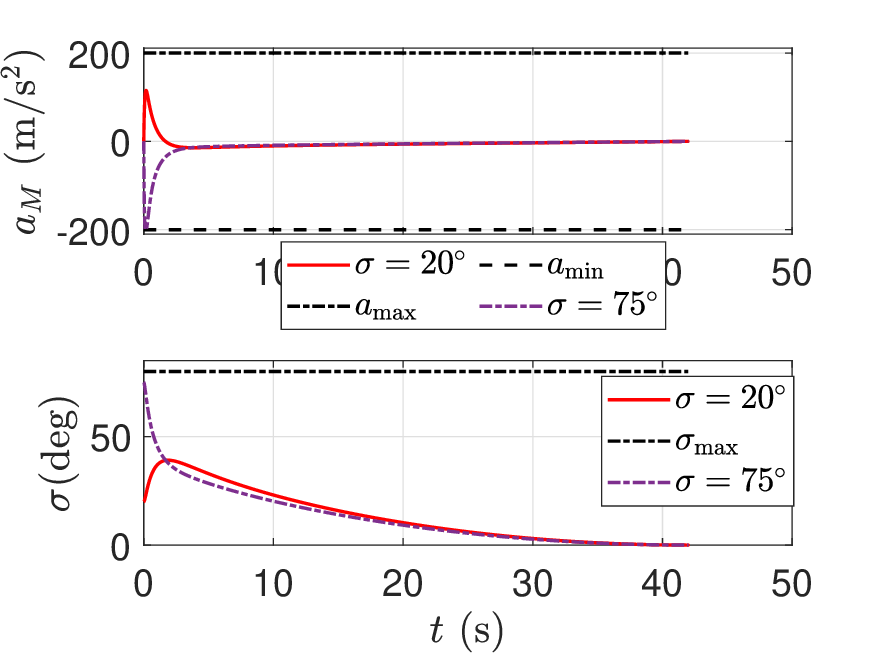}
\caption{Lateral acceleration and $\sigma$.}
\label{fig:Time42DifferentSigma_AMSigma}
\end{subfigure}
\caption{Target interception for $t_d = 42$s with different $\sigma$.}
\label{fig:Case1_Time42DifferentSigma}
\end{figure}
We first evaluate the performance with two different initial heading angles. The initial heading angles are set to $\gamma_M = 20^\circ$ and $\gamma_M = 75^\circ$, respectively, while the maximum allowable field-of-view (FOV) bound is taken as $\sigma_{\max} = 80^\circ$. The maximum lateral acceleration of the interceptor is constrained to $a_{\max}$. The simulation is performed for a desired impact time of $t_d = 42\,$s, and the results are presented in \Cref{fig:Case1_Time42DifferentSigma}.
From \Cref{fig:Time42DifferentSigma_XmYm}, we observe that the interceptor successfully intercepts the target at the desired impact time for both initial heading angles. This confirms that the proposed guidance law is independent of the interceptor's initial heading angle, which is a key feature of the design. 
As seen in \Cref{fig:Time42DifferentSigma_Time}, the time-to-go error nullifies to zero well before the actual interception, indicating that the time-to-go converges to its desired value at the same rate for both initial heading angles. This demonstrates that the guidance law achieves the desired impact time, regardless of the initial heading.

Additionally, from \Cref{fig:Time42DifferentSigma_AMSigma}, we observe that both the interceptor’s lateral acceleration and heading angle converge to zero in the neighborhood of the time of interception. During the transient phase (as shown in \Cref{fig:Time42DifferentSigma_AMSigma}), the lateral acceleration demand is higher. However, once the transient phase ends, that is, when the time-to-go error vanishes, the control demand decreases significantly and eventually converges to zero. This reduction in control effort is an important feature of the guidance law, as it ensures that the interceptor does not require excessive actuator resources.
Moreover, the heading angle error (or lead angle) also converges to zero well before the interception, which indicates that the interceptor aligns with the line of sight (LOS) well in advance of impact, striking the target along the LOS. Throughout the engagement, the interceptor always respects the imposed input and field-of-view constraints, as shown in \Cref{fig:Time42DifferentSigma_AMSigma}.
A closer inspection of \Cref{fig:Time42DifferentSigma_AMC} reveals that, for the case with $\gamma_M = 75^\circ$, the commanded lateral acceleration exceeds the physical actuator bounds. However, the design ensures that the interceptor’s actual lateral acceleration, $a_M$, never violates the given bounds. This reinforces the claim that the proposed guidance strategy guarantees interception at the desired impact time while strictly adhering to the physical limitations of the interceptor.
\subsection{Various impact times for interceptor’s fixed heading angle error}
\begin{figure*}[!ht]
\centering
\begin{subfigure}{0.33\linewidth}
	\centering
\includegraphics[width=\linewidth]{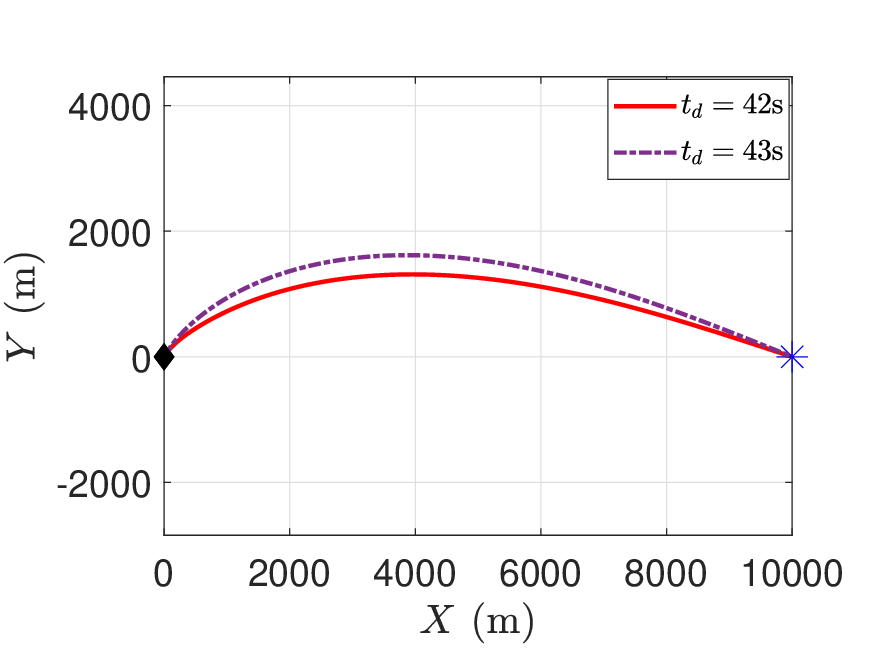}
    \caption{Trajectory.}
    \label{fig:Sigma60diffImpTime_XmYm}
\end{subfigure}%
\begin{subfigure}{0.33\linewidth}
	\centering
\includegraphics[width=\linewidth]{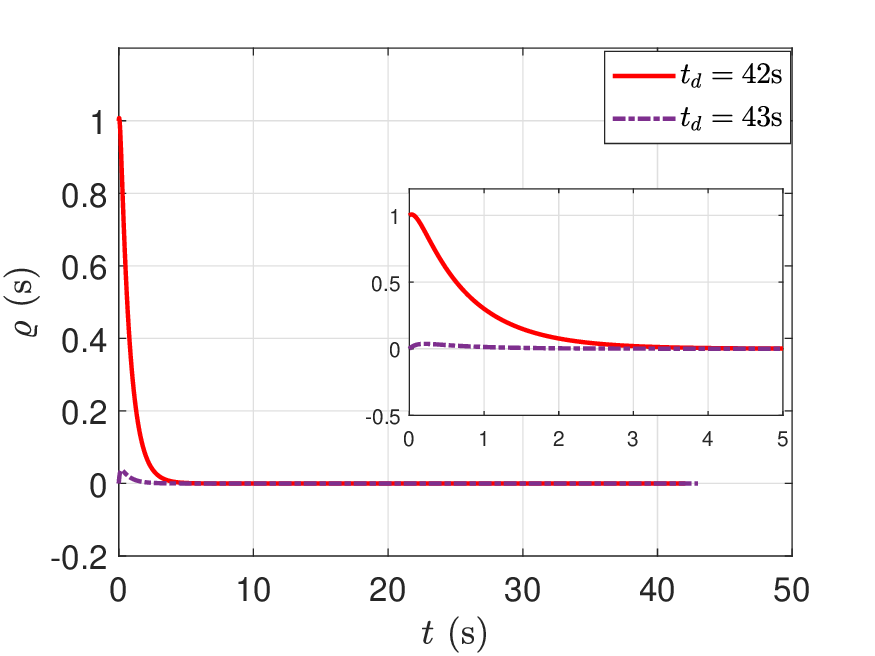}
    \caption{Impact time error.}
    \label{fig:Sigma60diffImpTime_err}
\end{subfigure}%
\begin{subfigure}{0.33\linewidth}
	\centering
\includegraphics[width=\linewidth]{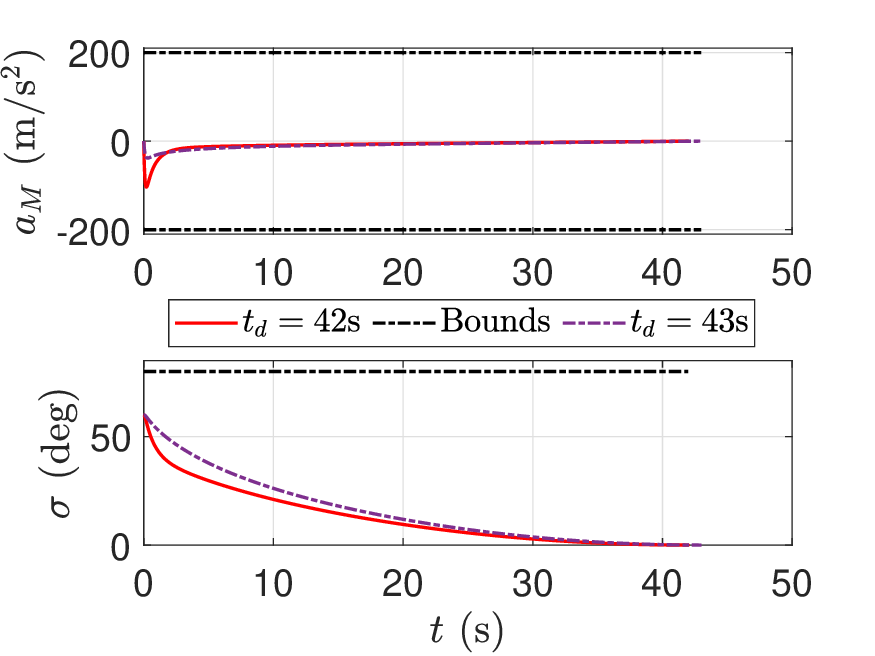}
    \caption{Lateral acceleration and $\sigma$.}
\label{fig:Sigma60diffImpTime_AMSigma}
\end{subfigure}
\caption{Target interception for $\sigma = 60^\circ$ with different $t_d$.}
\label{fig:Case2_Sigma60diffImpTime}
\end{figure*} 
\begin{figure*}[!ht]
\centering
\begin{subfigure}{0.4\linewidth}
	\centering
\includegraphics[width=\linewidth]{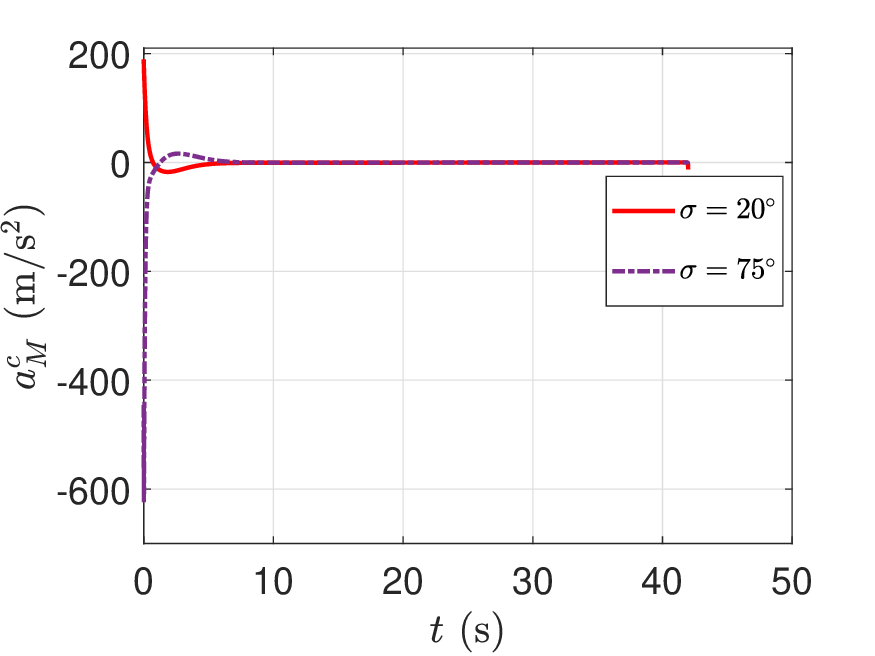}
    \caption{$t_d = 42$\,s}
    \label{fig:Time42DifferentSigma_AMC}
\end{subfigure}%
\begin{subfigure}{0.4\linewidth}
	\centering
\includegraphics[width=\linewidth]{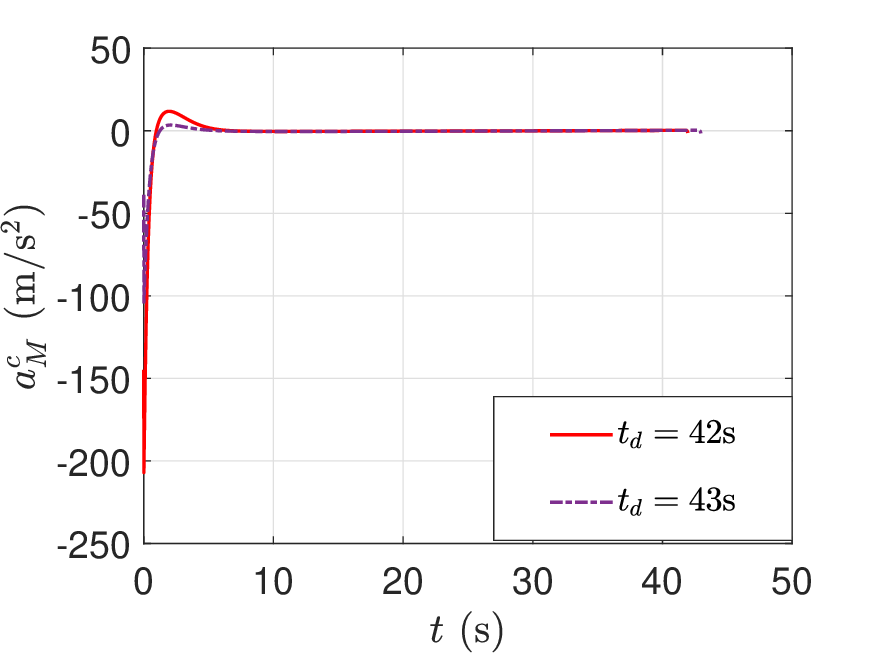}
    \caption{$\sigma = 60^\circ$.}
    \label{fig:Sigma60diffImpTime_AMC}
\end{subfigure}
\caption{Commanded lateral acceleration.}
\label{fig:AMC}   
\end{figure*}
Next, we validate the efficacy of the proposed guidance strategy by testing it under different impact time scenarios. The initial heading angle is set to $\gamma_M = 60^\circ$, with the maximum allowable field-of-view (FOV) bound $\sigma_{\max} = 80^\circ$. Simulations are carried out for two desired impact times: $t_d = 42\,\text{s}$ and $t_d = 43\,\text{s}$, while keeping all other parameters unchanged. The results are shown in \Cref{fig:Case2_Sigma60diffImpTime}.
From \Cref{fig:Sigma60diffImpTime_XmYm}, we observe that the interceptor successfully intercepts the target for both impact times. However, as expected, the interceptor follows a longer trajectory to intercept the target at $t_d = 43\,\text{s}$ compared to $t_d = 42 \, \text{s}$, which reflects the additional time available for interception. 

The time-to-go error is nullified well before the actual interception in both cases, as shown in \Cref{fig:Sigma60diffImpTime_err}. This confirms that the guidance law consistently drives the system towards the desired impact time, irrespective of the slight variation in the desired impact times.
As seen in \Cref{fig:Sigma60diffImpTime_AMSigma}, both the lateral acceleration and the heading angle error converge to zero for both impact time scenarios. Importantly, throughout the engagement, the interceptor’s lateral acceleration and lead angle never exceed their respective bounds, as illustrated in the same figure.
Moreover, \Cref{fig:Sigma60diffImpTime_AMC} highlights an interesting feature: for the $t_d = 42\, \text{s}$ case, the commanded lateral acceleration, $a_M^c$, exceeds the physical bound of $-20\,g$. However, the interceptor’s actual lateral acceleration, $a_M$, remains within the allowable bounds, as shown in \Cref{fig:Sigma60diffImpTime_AMSigma}. This demonstrates that the guidance law effectively respects the physical limitations of the interceptor while achieving the desired impact time.
\begin{figure*}[!ht]
\centering
\begin{subfigure}{0.33\linewidth}
	\centering
\includegraphics[width=\linewidth]{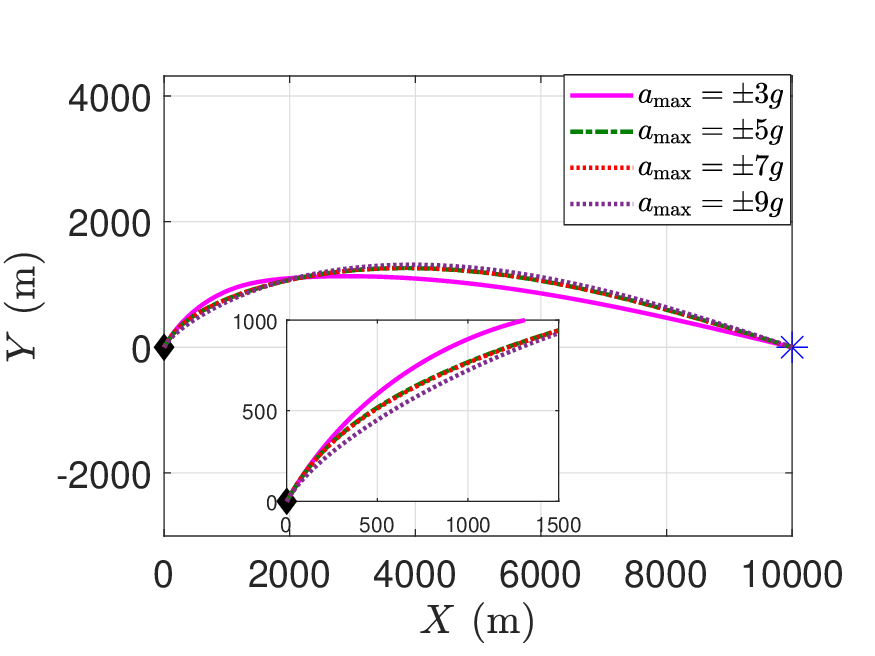}
    \caption{Trajectory.}
    \label{fig:DiffA_maxXmYm}
\end{subfigure}%
\begin{subfigure}{0.33\linewidth}
	\centering
\includegraphics[width=\linewidth]{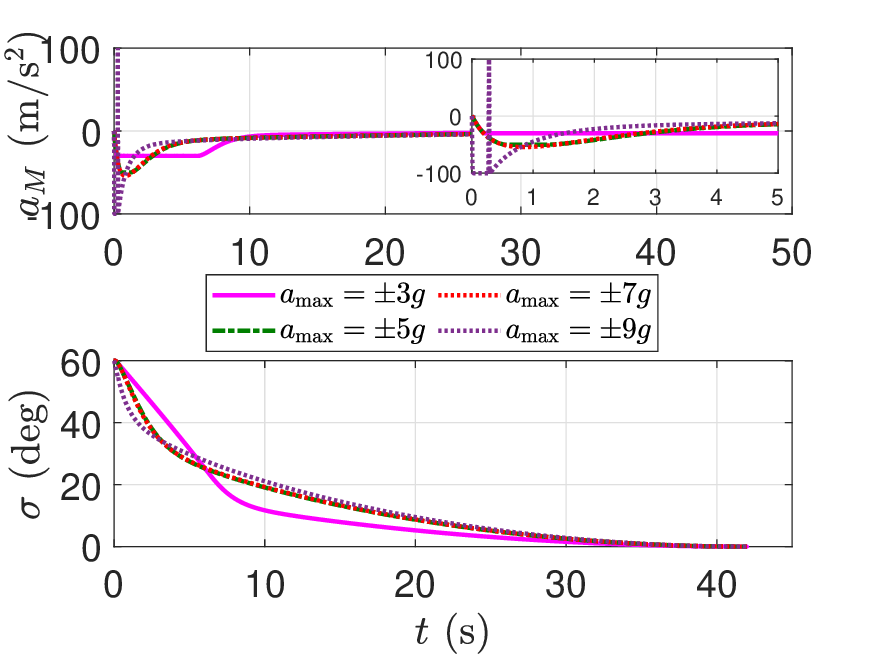}
    \caption{Lateral acceleration and $\sigma$.}
    \label{fig:DiffA_maxAMSigma}
\end{subfigure}%
\begin{subfigure}{0.33\linewidth}
	\centering
\includegraphics[width=\linewidth]{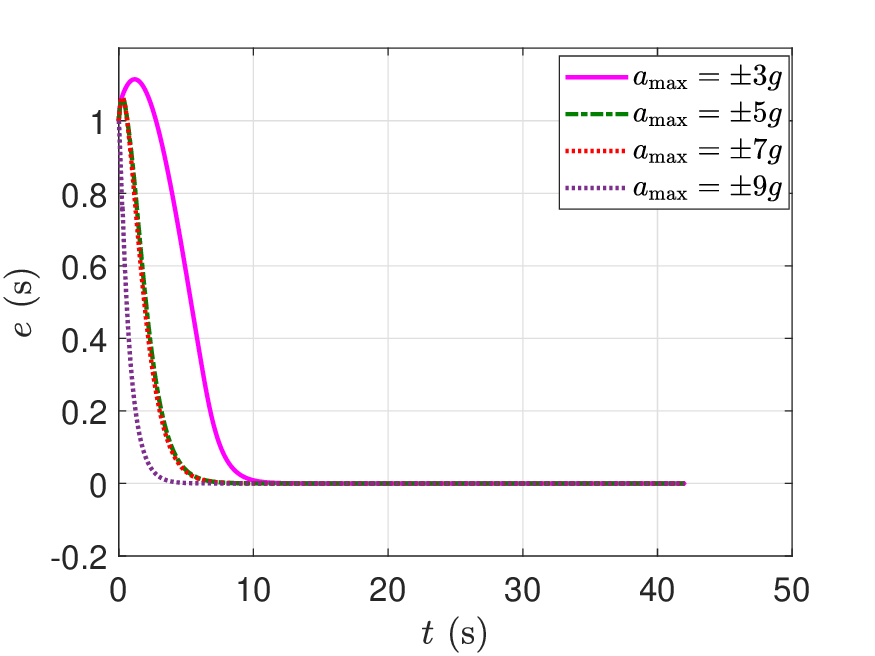}
    \caption{Error profiles.}
    \label{fig:DiffA_max_Err}
\end{subfigure}
\caption{Target interception for different $a_{\max}$.}
\label{fig:Case3_Different A_max}
\end{figure*}
\subsection{Performance with different $a_{\max}$ }
To further substantiate the efficacy of the proposed guidance law, simulations were performed for different values of the maximum allowable lateral acceleration, $a_{\max}$, while keeping the desired impact time fixed at $t_d = 42 \text{s}$ and the initial heading angle at $\sigma = 60^\circ$. The values of $a_{\max}$ tested were $\pm 3 g$, $\pm 5 g$, $\pm 7 g$, and $ \pm 9 g $. The simulation results are shown in \Cref{fig:Case3_Different A_max}.
As seen in \Cref{fig:DiffA_maxXmYm}, the proposed guidance law successfully intercepts the target for all tested acceleration bounds, demonstrating its robustness to varying control bounds. However, it is observed that the time-to-go error converges more slowly for smaller values of $a_{\max}$, as depicted in \Cref{fig:DiffA_max_Err}. This behavior is expected since lower lateral acceleration bounds constrain the interceptor's maneuvering capability, leading to a longer engagement time to reach the desired impact time.
Additionally, \Cref{fig:DiffA_maxAMSigma} illustrates the lateral acceleration demand, $a_M$, for each value of $ a_{\max}$. For the simulation parameters used, the lateral acceleration reaches its minimum value at $a_{\max} = \pm 3 \, g$, but it never exceeds the specified bounds. This confirms that the guidance law respects the physical constraints of the interceptor, even under reduced acceleration bounds.
\subsection{Comparison with existing guidance strategies}
\begin{figure*}[!ht]
		\centering
		\begin{subfigure}{0.33\linewidth}
			\includegraphics[width=\linewidth]{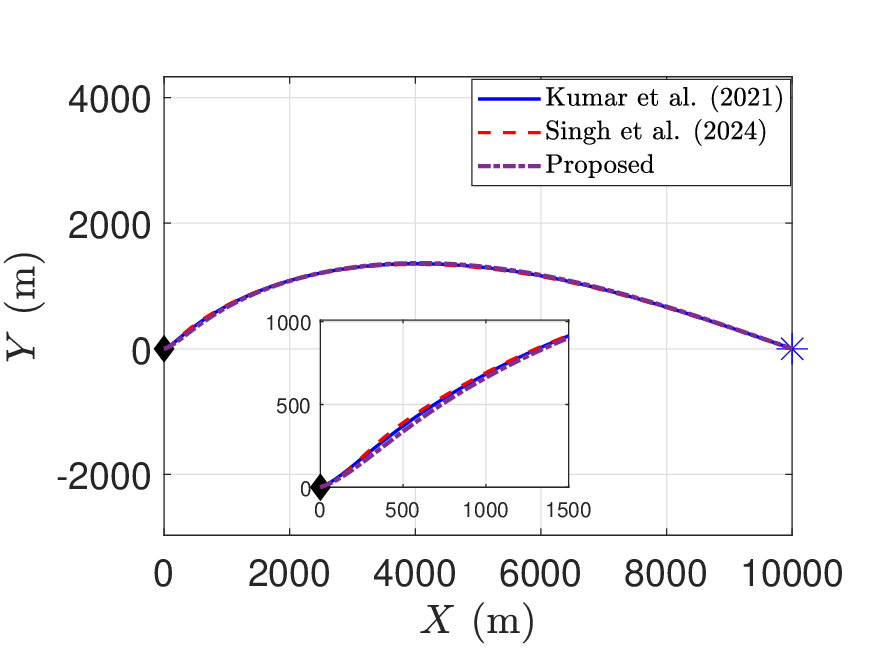}
			\caption{Trajectory.}\label{fig:comp_sim_trajectory}
		\end{subfigure}%
		\begin{subfigure}{0.33\linewidth}
			\includegraphics[width=\linewidth]{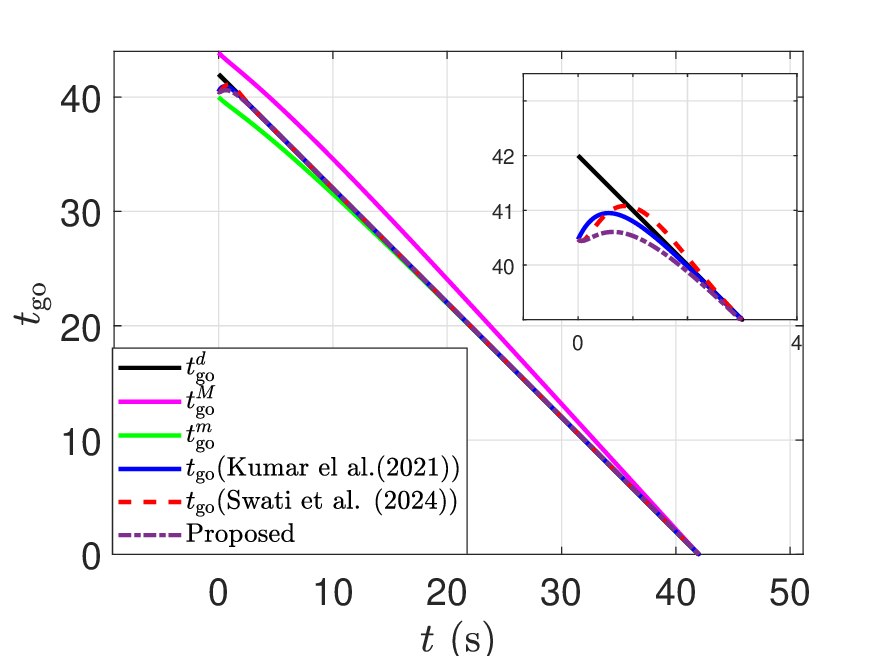}
			\caption{Time-to-go estimates.}\label{fig:comp_sim_Time}
		\end{subfigure}%
		\begin{subfigure}{0.33\linewidth}
			\includegraphics[width=\linewidth]{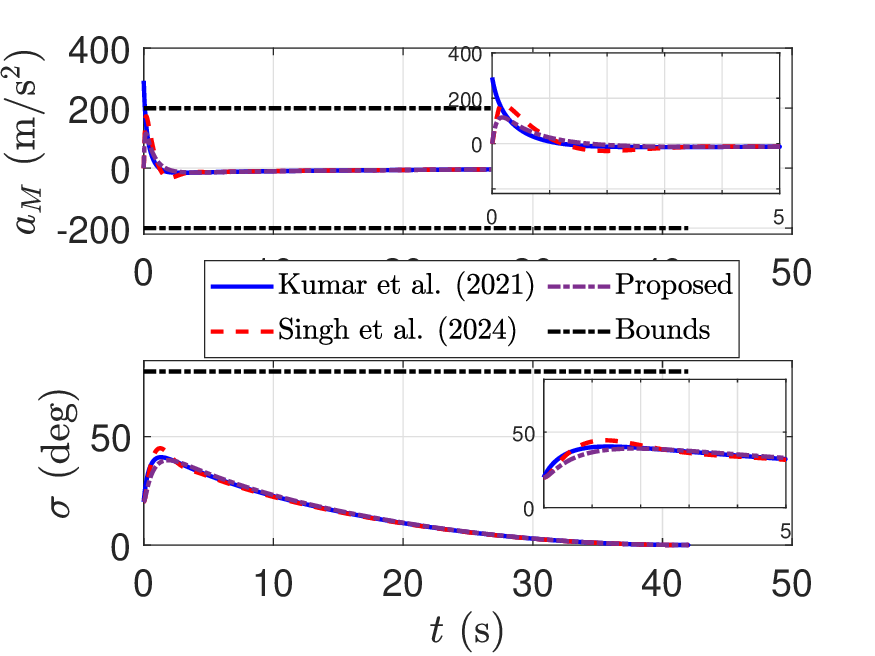}
			\caption{Lateral acceleration and $\sigma$.}\label{fig:comp_sim_AmSigma}
		\end{subfigure}
		\caption{Relative performance with existing guidance strategies.}
		\label{fig:comp_sim}
	\end{figure*}

To further validate the effectiveness of the proposed guidance law, we performed simulations to compare its performance against the results from \cite{kumar2021three} and \cite{doi:10.2514/1.G007770}. For these comparisons, we used the same simulation parameters as in previous sections. Notably, the strategy in \cite{kumar2021three} did not account for the physical bounds of the actuator in the guidance design, while the approach in \cite{doi:10.2514/1.G007770} considered a bound in the design but did not provide exact bounds for the lateral acceleration.
For consistency, the desired impact time was set to $t_d = 42\, \text{s}$, with the initial lead angle $\sigma = 60^\circ$. The performance of the three guidance strategies is depicted in \Cref{fig:comp_sim}. In all cases, a successful target interception is achieved. However, to further compare the performance of the strategies, we introduce the net control effort as a metric, defined as:
$\int_{0}^{t_{\rm imp}} {a_M}^2 dt$,
where $a_M$ is the lateral acceleration of the interceptor, and $t_{\rm imp}$ is the interception time. This metric quantifies the total control effort required to intercept the target. The comparison results for different impact times and initial heading angles are summarized in \Cref{tab:Tab1}, where the performance of the various engagement scenarios is presented.
In \textbf{Case 1}, we evaluated the control effort for different impact times, setting the maximum allowable field-of-view (FOV) bound $\sigma_{\rm max} = 80^\circ$ and choosing an initial heading angle of $60^\circ$. In \textbf{Case 2}, we fixed the impact time at $t_d = 42\, \text{s}$ and assessed the control effort for various initial heading angles. The results show that the proposed guidance strategy consistently outperforms the existing strategies in terms of the net control effort required for interception.
\begin{table}[!ht]
\centering
\caption{Total integral control effort for various engagement parameters.}\label{tab:Tab1}
\begin{tabular}{cccccccc}
\hline
\hline
\multirow{2}{*}{Methods} & \multicolumn{3}{c}{ Case 1} &                                                    & \multicolumn{3}{c}{\qquad  Case 2}                                                    \\ \cline{2-4} \cline{6-8}
                  & \multicolumn{1}{c}{$41$\,s} & \multicolumn{1}{c}{$42$\,s}  &$43$\,s  && \multicolumn{1}{c}{$20^\circ$} & \multicolumn{1}{c}{$40^\circ$}  & $60^\circ$ \\ \hline
    Proposed               &\multicolumn{1}{l}{$17953.9$} & \multicolumn{1}{c}{$9011.1$} & $5272.6$ && \multicolumn{1}{c}{$8037.1$} & \multicolumn{1}{c}{$2336$} & $9011$ \\ \hline
    \cite{doi:10.2514/1.G007770}              & \multicolumn{1}{c}{$23266$} & \multicolumn{1}{c}{$12115$}  & $5476.1$  && \multicolumn{1}{c}{$16101.8$} & \multicolumn{1}{c}{$2519.9$}  & $12115$ \\ \hline
    \cite{kumar2021three}              & \multicolumn{1}{c}{$528687$} & \multicolumn{1}{c}{$326294$} & $52212.5$  && \multicolumn{1}{c}{$1584556.5$} & \multicolumn{1}{c}{$250705.7$}& $1063820$ \\ \hline
    \hline
\end{tabular}
\end{table}
\subsection{Performance with larger impact time}
\begin{figure*}[!ht]
\begin{subfigure}{0.33\linewidth}
	\centering
\includegraphics[width=\linewidth]{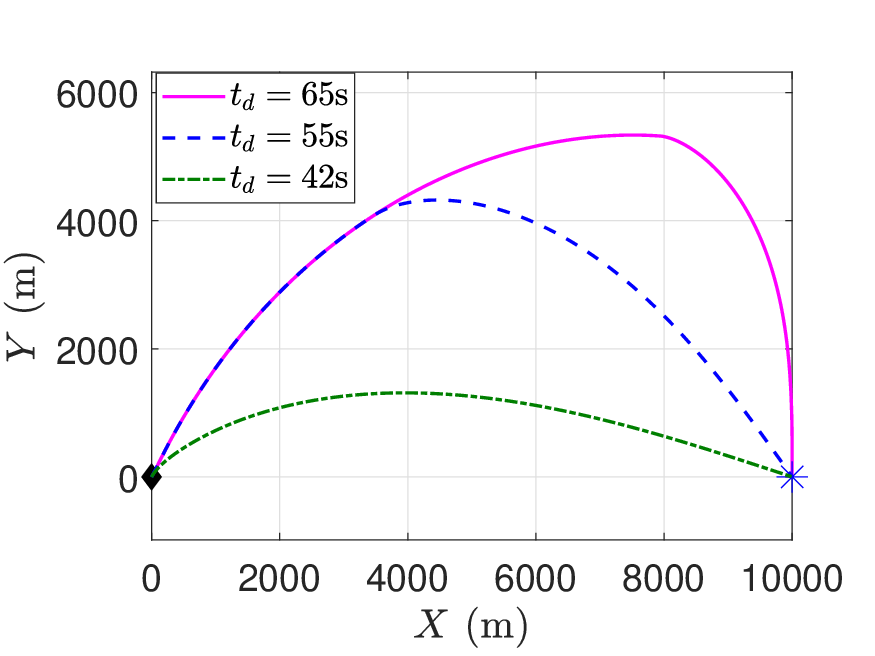}
    \caption{Trajectory.}
    \label{fig:LargeImpTime_XmYm}
\end{subfigure}%
\begin{subfigure}{0.33\linewidth}
	\centering
\includegraphics[width=\linewidth]{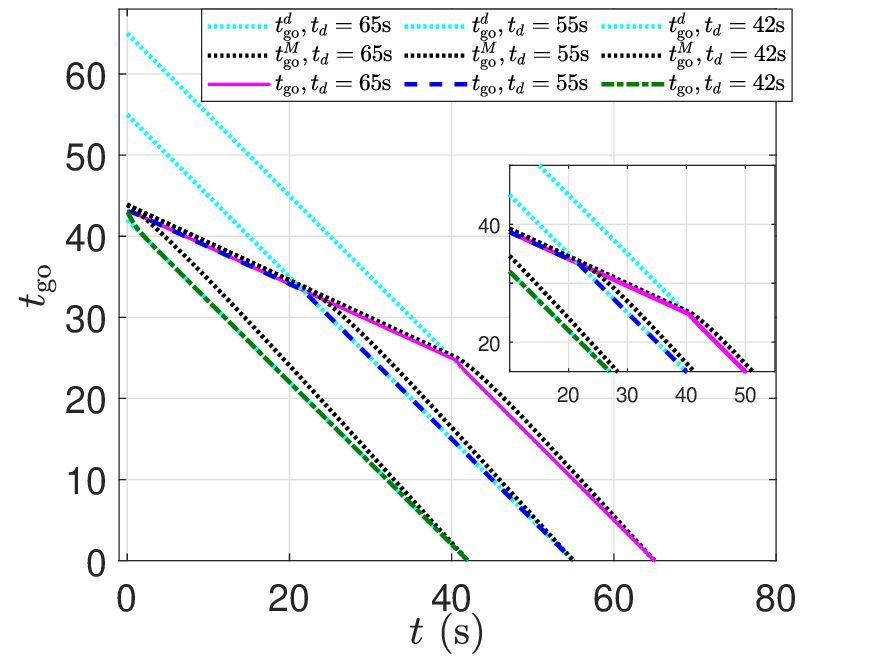}
    \caption{Time-to-go estimates.}
    \label{fig:LargeImpTime_Time}
\end{subfigure}%
\begin{subfigure}{0.33\linewidth}
	\centering
\includegraphics[width=\linewidth]{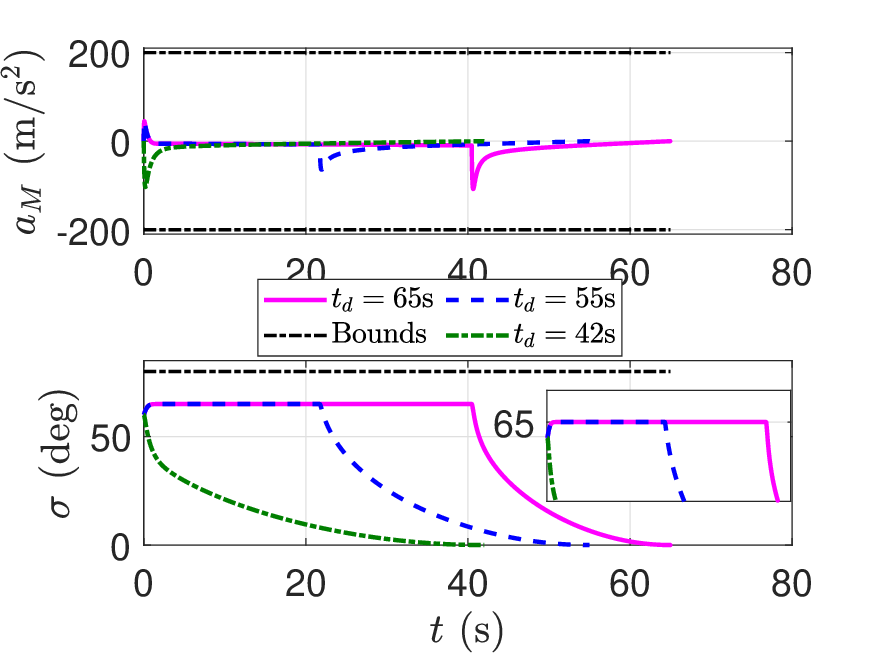}
    \caption{Lateral acceleration and $\sigma$.}
    \label{fig:LargeImpTime_AMSigma}
\end{subfigure}
\begin{subfigure}{0.33\linewidth}
	\centering
\includegraphics[width=\linewidth]{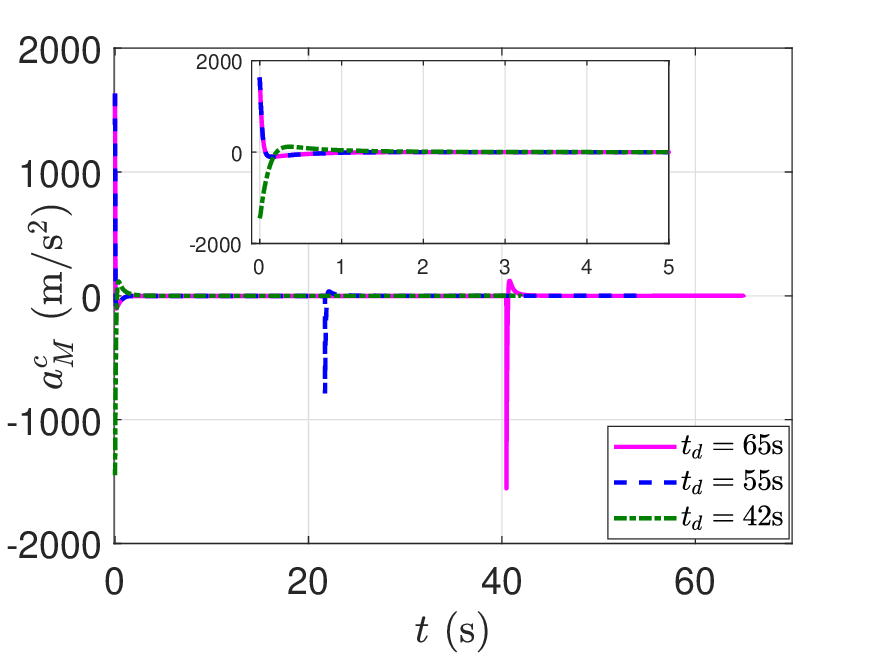}
    \caption{Commanded lateral acceleration.}
    \label{fig:LargeImpTime_AMC}
\end{subfigure}%
\begin{subfigure}{0.33\linewidth}
	\centering
\includegraphics[width=\linewidth]{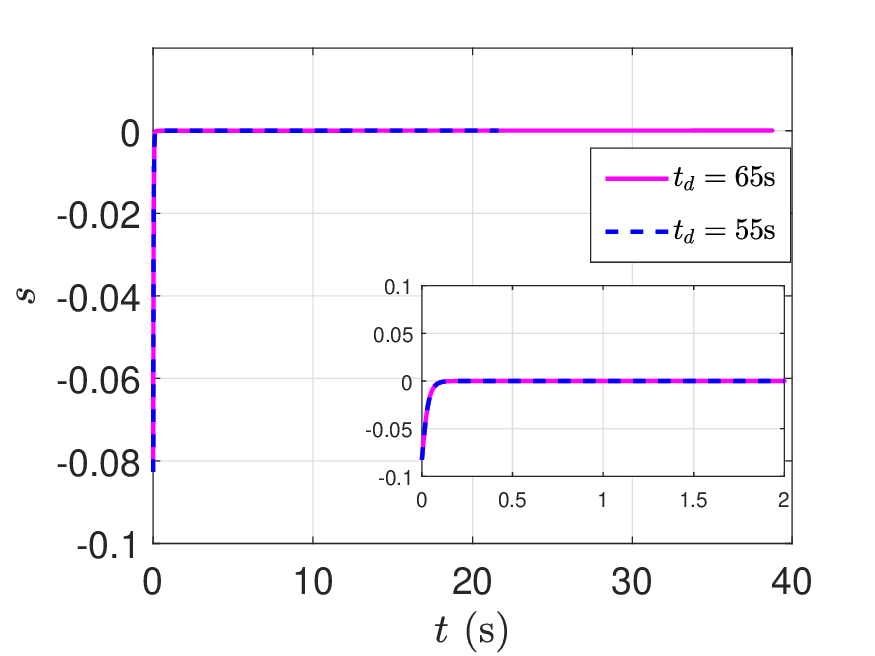}
    \caption{Sliding surface.}
    \label{fig:LargeImpTime3S_SS}
\end{subfigure}%
\begin{subfigure}{0.33\linewidth}
	\centering
\includegraphics[width=\linewidth]{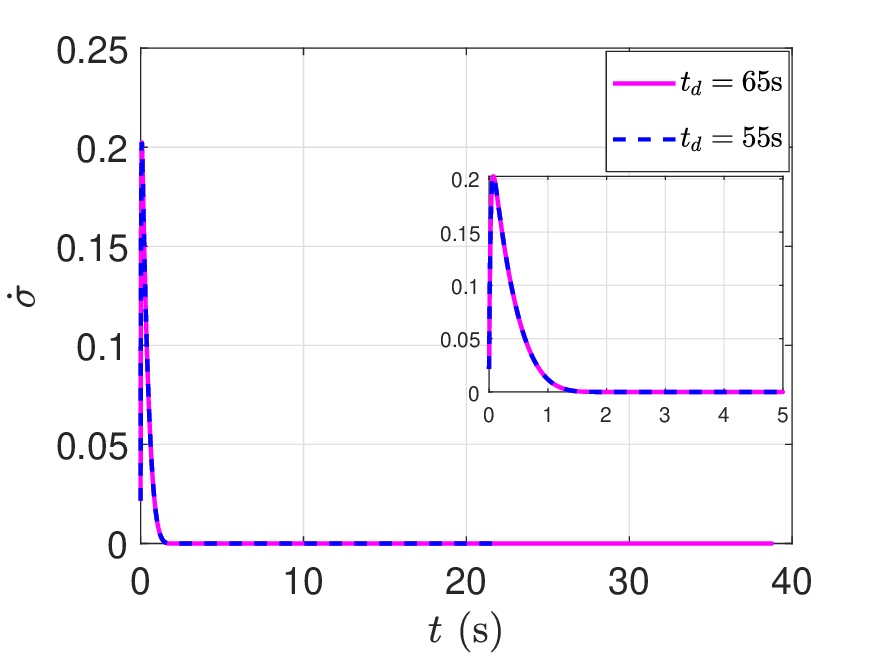}
    \caption{Heading angle error rate, $\dot{\sigma}$.}
    \label{fig:LargeImpTime3S_SigmaDot}
\end{subfigure}
\caption{Target interception for larger impact times.}
\label{fig:Case4_Large Impact Time}
\end{figure*}

The guidance strategy proposed in \Cref{sec3B} performs effectively when the desired impact time is within its allowable range. To assess its performance for impact times exceeding the maximum achievable value $t_{\rm go}^M$, we apply the deviated pursuit strategy outlined in \Cref{sec3C}. This approach first achieves a larger impact time and then transitions to the strategy in \Cref{sec3B}. Simulation results for this approach are shown in \Cref{fig:Case4_Large Impact Time}.
We tested three impact times: 42,s, 55,s, and 65\,s, with a desired heading angle $\sigma_d = 65^\circ$. As shown in \Cref{fig:LargeImpTime_XmYm}, the interceptor successfully intercepted the target in each case. The corresponding switching times $t_1$ for these impact times were 0 s, 24.39 s, and 44.63 s, as confirmed in \Cref{fig:LargeImpTime_Time}. At these switching points, the desired time-to-go $(t_{\rm go}^d$) matches the maximum achievable time-to-go $(t_{\rm go}^M$), ensuring that both FOV and input constraints are satisfied.
Initially, the interceptor increases its heading error to achieve $\sigma_d$ and then follows a deviated pursuit with a constant heading angle $(\dot{\sigma} = 0$) until $t_1$. Afterward, the heading error decreases, converging to zero, as shown in \Cref{fig:LargeImpTime_AMSigma}. Sliding mode is active until $t_1$ for the 55 s and 65 s impact times, as seen in \Cref{fig:LargeImpTime3S_SS}. At $t_1$, the strategy transition causes jumps in both lateral acceleration and commanded acceleration, as shown in \Cref{fig:LargeImpTime_AMSigma} and \Cref{fig:LargeImpTime_AMC}, respectively, however, the lateral acceleration remains within the allowed bounds.

\subsection{Performance with Second-order Autopilot}
\begin{figure}
    \centering
    \includegraphics[width=\linewidth]{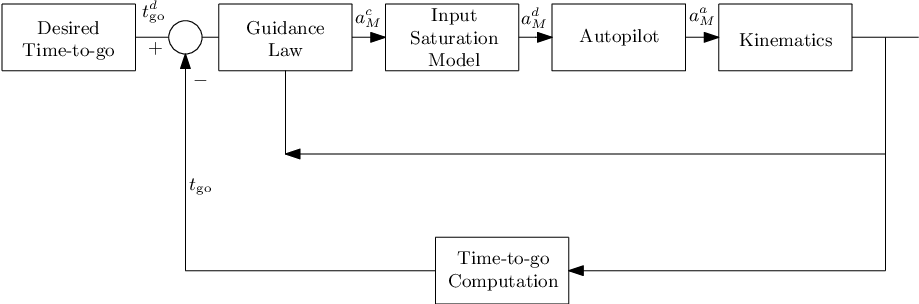}
    \caption{System in the presence of Autopilot}
    \label{fig:Autopilot_Block}
\end{figure}

Next, we demonstrate the performance of our guidance strategy in the presence of an autopilot in the simulation. It has
been shown in \citep{doi:10.2514/1.34042} that a 13\textsuperscript{th} order autopilot can be truncated to two first-order
ones via some balanced realization. Hence, it is pragmatic to capture the higher-order dynamical phenomenon using a lower-order approximation, resulting in the traceability of physical realization, while preserving the behaviors that may arise due to autopilot dynamics. The schematic of the proposed guidance strategy in the presence of autopilot has been shown in the \Cref{fig:Autopilot_Block}. Let us consider the autopilot dynamics as 
\begin{equation}\label{eq:Autopilot}
    \dfrac{a_{M}^a}{a_{M}^d} = \dfrac{1}{s\tau+1},
\end{equation}
where $a_{M}^a$ and $a_{M}^d$ denote the achieved and the desired lateral accelerations, respectively. Note that here, the term $a_{M}^d$ is nothing but the output of the input model. The dynamics presented in \Cref{eq:Autopilot} is of first-order with $\tau$ as the time constant. Similarly,
the 13\textsuperscript{th} order autopilot approximated as a second-order dynamics \citep{doi:10.2514/1.34042} can be written as
\begin{equation}
    \dfrac{a_{M}^a}{a_{M}^d} = \dfrac{1}{(0.56s+1)(0.1s+1)}.
\end{equation}

\begin{figure*}[!ht]
\begin{center}
   \begin{subfigure}{0.33\linewidth}
	\centering
\includegraphics[width=\linewidth]{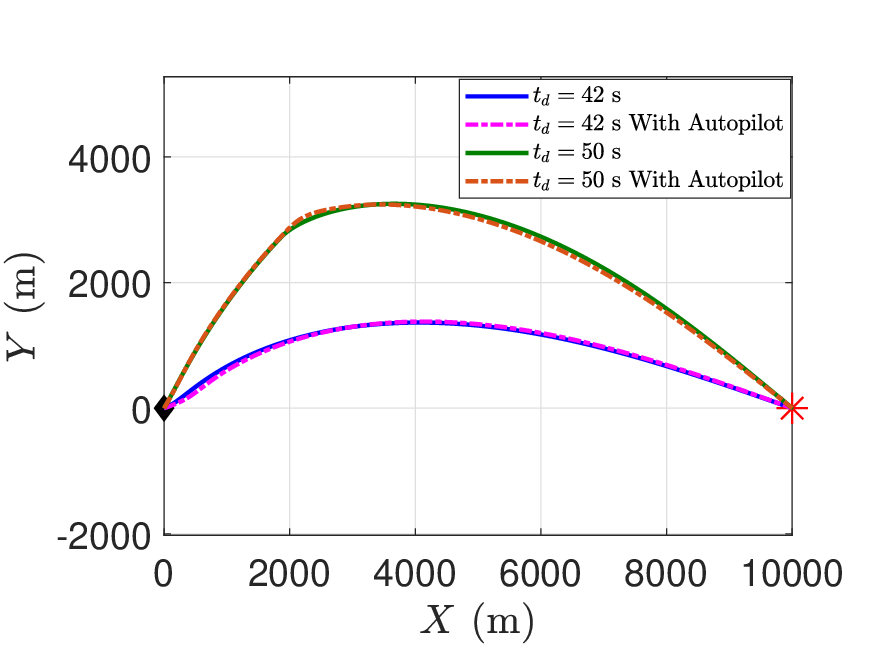}
    \caption{Trajectory.}
    \label{fig:Autopilot_XmYm}
\end{subfigure}%
\begin{subfigure}{0.33\linewidth}
	\centering
\includegraphics[width=\linewidth]{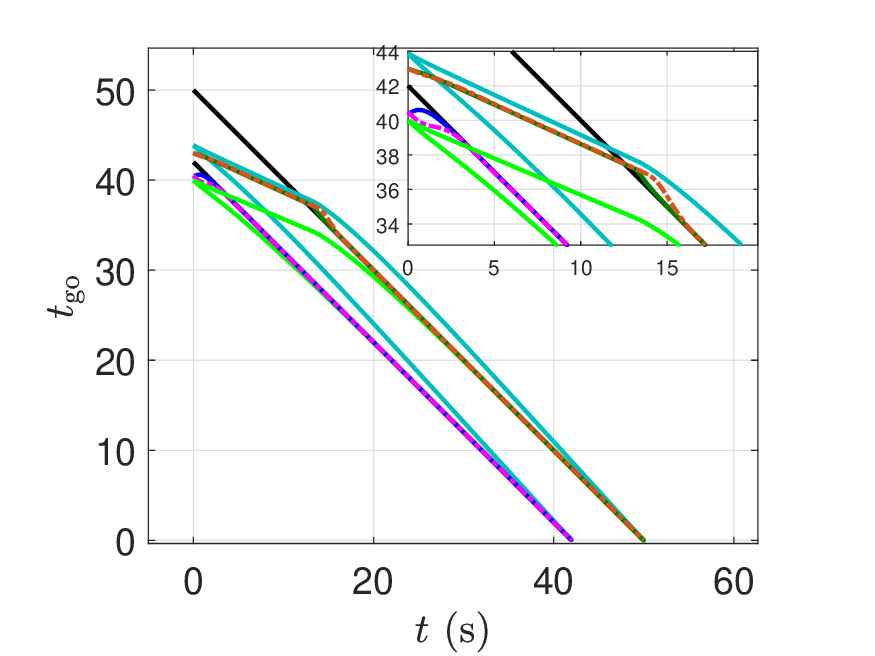}
    \caption{Time-to-go estimates.}
    \label{fig:Autopilot_Tgo}
\end{subfigure}%
\begin{subfigure}{0.33\linewidth}
	\centering
\includegraphics[width=\linewidth]{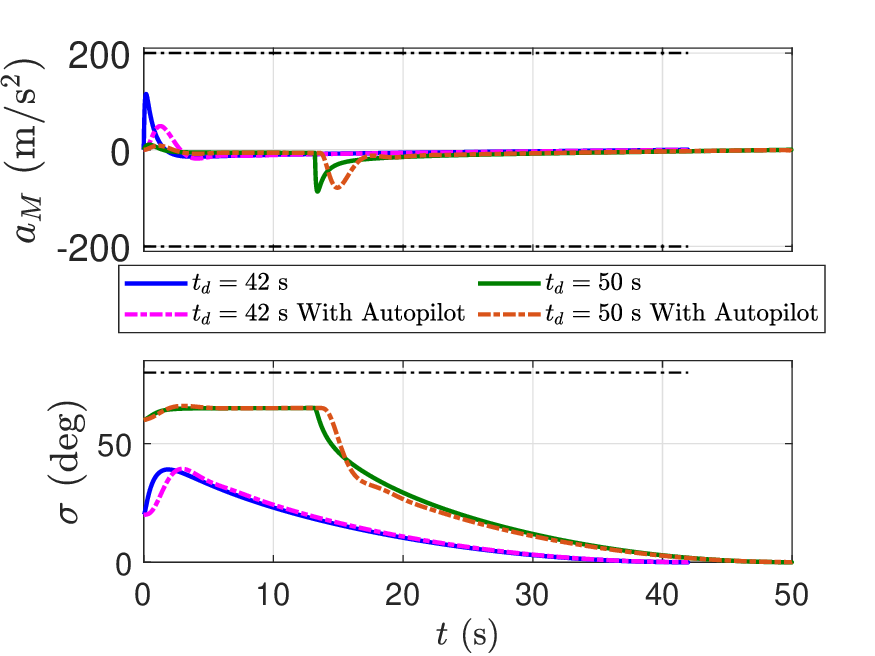}
    \caption{$\sigma$ and Lateral Acceleration.}
    \label{fig:Autopilot_Accele}
\end{subfigure}
\end{center}
\caption{Target interception for different impact times.}
\label{fig:Case5_Autopilot}   
\end{figure*}
\Cref{fig:Case5_Autopilot} depicts the performance of the proposed guidance strategy under the second-order autopilot as given in \eqref{eq:Autopilot}. For this case, interceptions at two different impact times, $t_d = 42$ s and $t_d = 50$ s, were performed with heading angle errors of $\sigma = 20^\circ$, $\sigma = 60^\circ$, and $\sigma_d = 65^\circ$. It is easy to verify from \Cref{fig:Case5_Autopilot} that although the lateral acceleration demands differ slightly due to the presence of lag, the target interception is still guaranteed at the desired impact time, considering both the actuator constraints.

\subsection{Performance with Measurement Noise}
\begin{figure*}[!ht]
\begin{center}
   \begin{subfigure}{0.33\linewidth}
	\centering
\includegraphics[width=\linewidth]{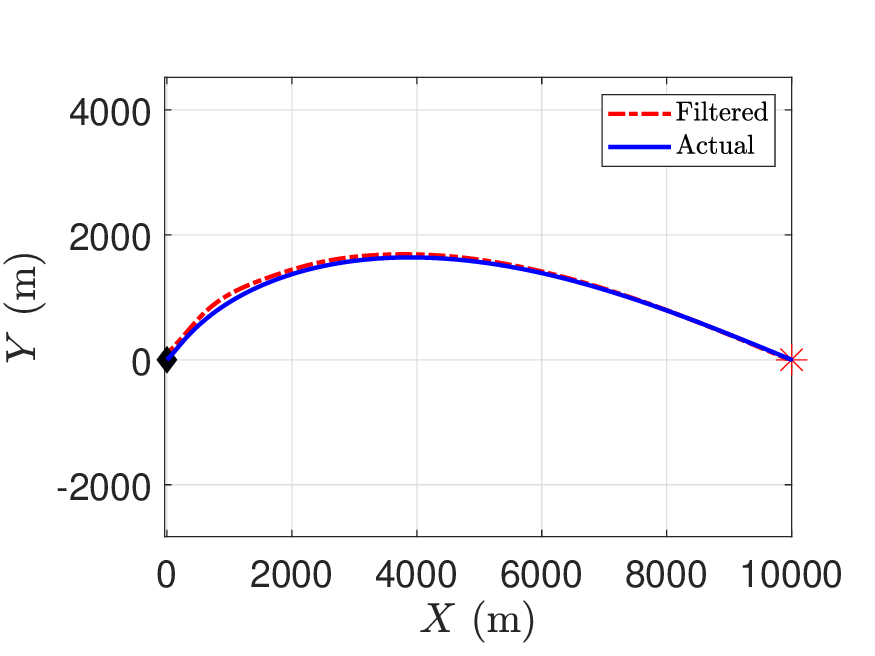}
    \caption{Trajectory.}
    \label{fig:XmYm_Noise}
\end{subfigure}%
\begin{subfigure}{0.33\linewidth}
	\centering
\includegraphics[width=\linewidth]{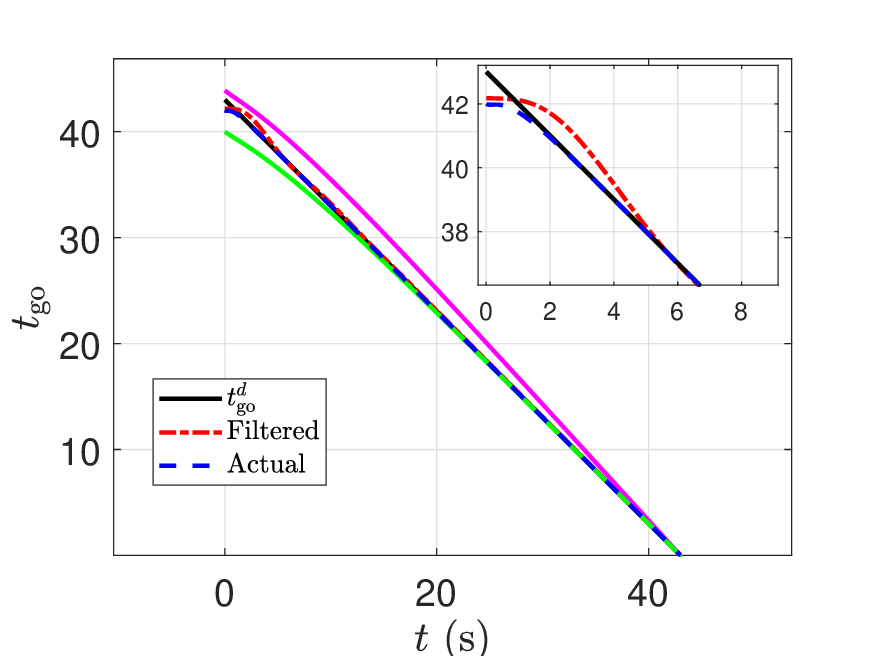}
    \caption{Time-to-go estimates.}
    \label{fig:Time_Noise}
\end{subfigure}%
\begin{subfigure}{0.33\linewidth}
	\centering
\includegraphics[width=\linewidth]{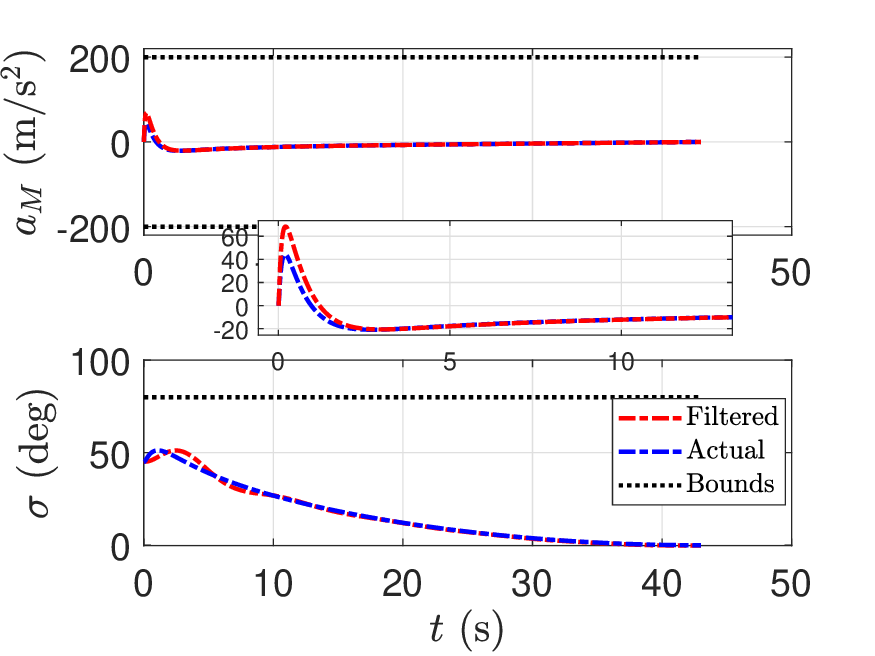}
    \caption{$\sigma$ and Lateral Acceleration.}
    \label{fig:AmSigma_Noise}
\end{subfigure}
\end{center}
\caption{Target interception under noisy measurements.}
\label{fig:Case6Noise}   
\end{figure*}

While the proposed strategy assumes perfect measurements, this is rarely the case in practice. Measurements obtained by onboard seekers are often noisy. This promotes the evaluation of the proposed guidance strategy under imperfect measurements. The LOS and heading angles are assumed to be corrupted by Gaussian noise with zero mean and a standard deviation of $15$ mrad. Range measurements are also considered imperfect, within $\pm 1\%$ of the actual value. Furthermore, the sensors are assumed to have a sampling rate of $100$ Hz. In order to investigate the performance under noise, we have employed an $\alpha$--$\beta$ filter, \cite{zarchan2012tactical}, which is a steady-state two-dimensional Kalman filter in essence, and can also be viewed as a simple state observer without requiring a detailed system model. The performance of the filter relies on the parameters $\alpha$--$\beta$. The relevant plots are shown in \Cref{fig:Case6Noise}. One may observe from \Cref{fig:XmYm_Noise} that when the estimation error converges, the interceptor follows a course similar to that in the absence of noise. A similar conclusion can be made for the time-to-go estimate in \Cref{fig:Time_Noise}. The behavior of lead angle, lateral accelerations components, and their rate are depicted in \Cref{fig:AmSigma_Noise}, from which we can draw similar conclusions.

\begin{figure*}[!ht]
\begin{center}
   \begin{subfigure}{0.33\linewidth}
	\centering
\includegraphics[width=\linewidth]{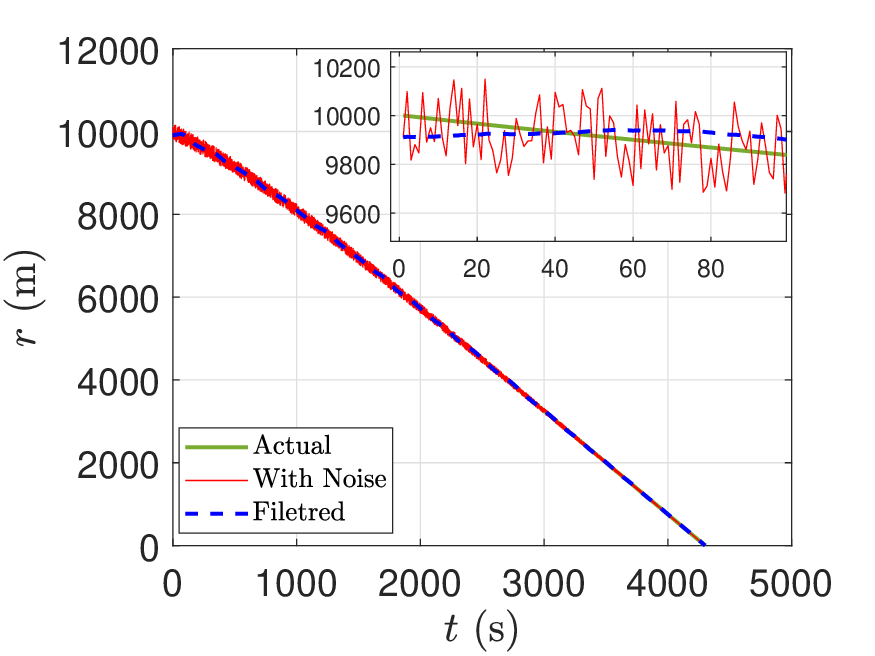}
    \caption{Range measurement.}
    \label{fig:R_Noise}
\end{subfigure}%
\begin{subfigure}{0.33\linewidth}
	\centering
\includegraphics[width=\linewidth]{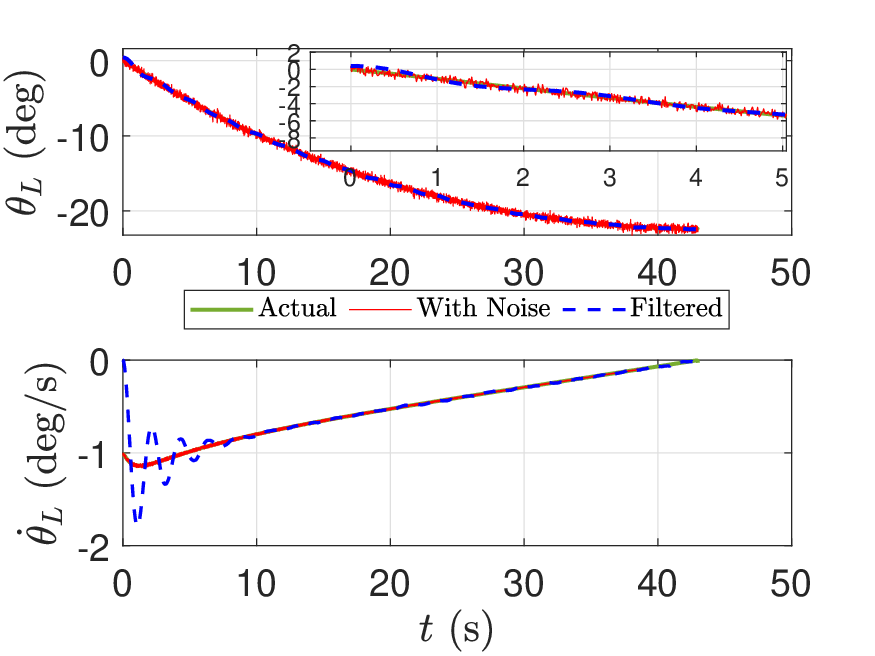}
    \caption{Measurements in LOS angle.}
    \label{fig:ThlThldot_Noise}
\end{subfigure}%
\begin{subfigure}{0.33\linewidth}
	\centering
\includegraphics[width=\linewidth]{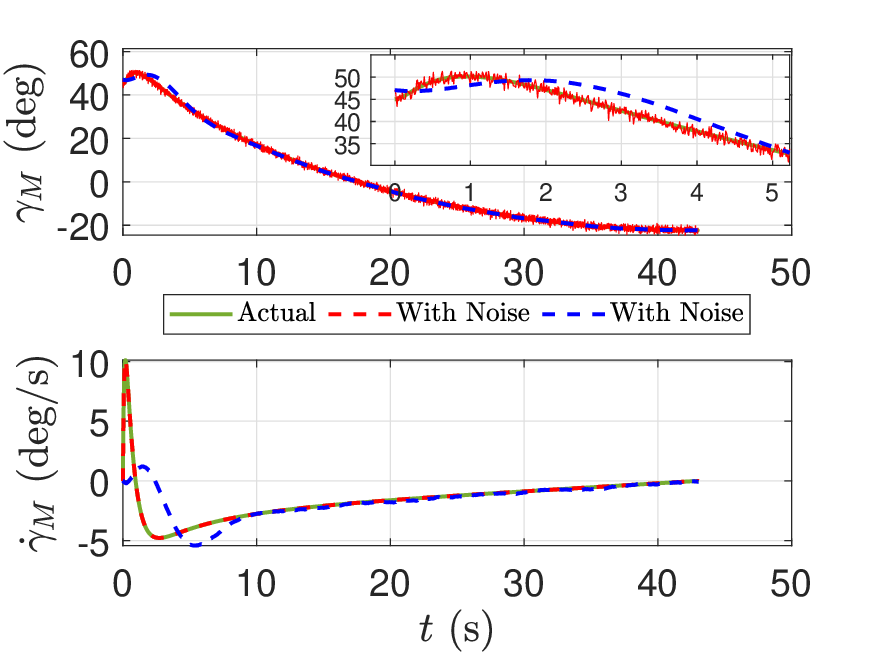}
    \caption{Measurements in flight path angle.}
    \label{fig:GamaGamadot_Noise}
\end{subfigure}
\end{center}
\caption{Actual, noisy, and filtered measurements.}
\label{fig:Case7Noisedata}   
\end{figure*}

The behaviors of angular and range measurements in the presence of noise are
shown in \Cref{fig:Case7Noisedata}. It is observed that after a transient period, the filtered values track the true values (actual values) despite the presence of noise. 
A zoomed-in view is provided in the corresponding sub-figures for clarity. These observations demonstrate the robustness and effectiveness of the proposed strategy under noisy conditions.

\section{Conclusions}\label{sec5}
In this paper, we have presented a novel nonlinear guidance law for planar engagement scenarios, ensuring precise target interception at a desired impact time while adhering to both field-of-view (FOV) and control input constraints. The proposed guidance strategy is particularly advantageous in nonlinear settings, remaining effective even in situations with large initial heading errors, which are common in real-world engagements. To address the physical limitations of the interceptor, we incorporated a smooth input saturation model to enforce lateral acceleration constraints without sacrificing performance.
Our results demonstrate that, despite these constraints, the proposed guidance law enables the interceptor to successfully meet the impact time requirements and achieve target interception. Specifically, both the heading angle error and lateral acceleration converge to zero as the interceptor reaches the target, ensuring a smooth and accurate interception trajectory. These findings highlight the robustness of the proposed approach, which effectively integrates the physical constraints without compromising the mission objective.
Additionally, we introduced a multi-stage strategy to extend the achievable range of impact times. In this approach, the interceptor initially increases its heading angle error to create the necessary conditions for a deviated pursuit course, during which it maintains a constant lead angle. Upon reaching a predefined switching instant, the interceptor adjusts its heading angle to smoothly converge toward the target, ultimately ensuring a successful interception. This multi-stage maneuver proves effective for handling larger impact times while maintaining stability and control.
Finally, we have shown that the proposed guidance strategy not only ensures successful target interception at larger impact times but also respects the physical constraints of the actuator and seeker. The results suggest that this approach can be applied to a wide range of real-world guidance systems, where practical limitations such as actuator capabilities and sensor field-of-view must be accounted for in the design.
\bibliography{aiaa}
\end{document}